\documentstyle[prd,aps,twocolumn,epsf,floats,amsmath,amssymb]{revtex} 
\newcommand{\ignore}[1]{}
\newcommand{\Ref}[5]{#1, #2 {\bf #3}, #4 (#5)}
\newcommand{\Refa}[4]{#1 {\bf #2}, #3 (#4)}
\newcommand{\Refb}[3]{{\bf #1}, #2 (#3)}
\newcommand{\PRD}{Phys.~Rev.~D}
\newcommand{\C}{{\mathbb C}}
\newcommand{\N}{{\mathbb N}}
\newcommand{\R}{{\mathbb R}}
\newcommand{\s}{{\mathbb S}}
\newcommand{\Z}{{\mathbb Z}}

\newcommand{\Mc}{{\mathcal M}}
\newcommand{\Qc}{{\mathcal Q}}
\newcommand{\Rc}{{\mathcal R}}

\newcommand{\Uc}{{\mathcal U}}
\newcommand{\Vc}{{\mathcal V}}
\newcommand{\Xc}{{\mathcal X}}
\newcommand{\Yc}{{\mathcal Y}}

\newcommand{\Qct}{{\widetilde{{\mathcal Q}}}}

\newcommand{\Uct}{{\widetilde{{\mathcal U}}}}
\newcommand{\Vct}{{\widetilde{{\mathcal V}}}}
\newcommand{\Xct}{{\widetilde{{\mathcal X}}}}
\newcommand{\Yct}{{\widetilde{{\mathcal Y}}}}

\newcommand{\etat}{{\tilde{\eta}}}
\newcommand{\gt}{{\tilde{g}}}
\newcommand{\hti}{{\tilde{h}}}
\newcommand{\Kt}{{\widetilde{K}}}

\newcommand{\Omt}{{\widetilde{\Omega}}}

\newcommand{\qt}{{\tilde{q}}}

\newcommand{\ut}{{\tilde{u}}}
\newcommand{\vt}{{\tilde{v}}}
\newcommand{\xt}{{\tilde{x}}}
\newcommand{\yt}{{\tilde{y}}}
\newcommand{\zt}{{\tilde{z}}}

\newcommand{\xf}{{\bf x}}
\newcommand{\xft}{{\tilde{\mathbf x}}}
\newcommand{\yf}{{\bf y}}
\newcommand{\yft}{{\tilde{\mathbf y}}}
\newcommand{\QR}{{_R}}
\newcommand{\QF}{{_F}}
\newcommand{\QL}{{_L}}
\newcommand{\QP}{{_P}}
\newcommand{\QI}{{_I}}
\newcommand{\QII}{{_{I\!I}}}
\newcommand{\ED}{\end{document}}

\begin{document}

\twocolumn[\hsize\textwidth\columnwidth\hsize\csname@twocolumnfalse%
\endcsname%

\hfill{Imperial/TP/98-99/38, UGVA-DPT-1999/05-1041} \\

\title{Quantum mechanical path integrals and thermal radiation in
static curved spacetimes}

\author{F.~Vendrell}  

\address{Blackett Laboratory, Imperial College, Prince Consort Road,
London SW7 2BZ, United Kingdom 
\\ and \\ 
D\'epartement de physique th\'eorique, Universit\'e de Gen\`eve, 
Quai E.~Ansermet 24, 1211 Geneve 4, Switzerland }

\date{June 3, 1999} 
 
\maketitle

\begin{abstract}

The propagator of a spinless particle is calculated from the quantum
mechanical path integral formalism in static curved spacetimes endowed
with event-horizons. A toy model, the Gui spacetime, and the 2D and 4D
Schwarzschild black holes are considered.  The role of the topology of
the coordinates configuration space is emphasised in this framework.
To cover entirely the above spacetimes with a single set of
coordinates, tortoise coordinates are extended to {\it complex}
values. It is shown that the homotopic properties of the complex
tortoise configuration space imply the thermal behaviour of the
propagator in these spacetimes. The propagator is calculated when end
points are located in identical or distinct spacetime regions
separated by one or several event-horizons.  Quantum evolution through
the event-horizons is shown to be unitary in the fifth variable.

\end{abstract}


]

\section{Introduction} \label{sec:intro}

Quantum mechanical path integrals \cite{FHS,Ma,Sch1,KL}, also called
first quantised path integrals, have been applied to some problems in
curved spacetimes \cite{BP}, such as to cosmological and black-holes
issues \cite{HH,TV,Co,Be,Me,OV}. A remarkable theoretical prediction in
semi-classical gravity is that of the thermal and quantum radiation of
black holes \cite{Ha,BD}. This result is recovered again in the
present paper within the formalism of quantum mechanical path
integrals. We shall consider some static spacetimes endowed with
event-horizons such as the two-dimensional (2D) and four-dimensional
(4D) Schwarzschild black-hole spacetimes, and a toy spacetime model,
the Gui spacetime \cite{Gui1,Gui2}.

\bigskip
An advantage of the path integral formalism over the canonical
approach of quantum field theory is that path integrals capture at
once both the local and global properties of the base space in which
there are calculated. They are thus quite useful in spaces endowed
with a non-trivial topology, since they allow one to exploit
relatively easily the homotopic properties of the space when computing
the propagator.

\smallskip
The circle is the simplest example of a space with a non-trivial
topology in which path integrals can be calculated \cite{Ma,Sch1}. In
this case, paths are catalogued into homotopic classes according to
their number of turns around the circle. This number is called the
winding number (it is positive when the path turns in the
anticlockwise direction, and negative when it turns in the clockwise
direction).  A path contribution to the path integral depends
essentially on the homotopic class to which it belongs, i.e.~its
contribution is a function of its winding number. One can decompose
the propagator as the sum of the individual contributions of the
homotopic classes over the winding number with respect to the
circle. The contributions of the homotopic classes can be calculated
in this case and shown to be equal to the free propagator. In more
complex spaces, path integrals are evaluated in a similar way. One
needs sometimes to define several winding numbers \cite{Isham}, but
the contributions of the homotopic classes may be quite difficult to
calculate, as in curved backgrounds.

\bigskip
The explicit dependence of path integrals on homotopic properties may
also be useful when physical or topological constraints are introduced
in a base space whose topology is {\it trivial}. In that case, one
usually prefers to work in a copy of the base space from which the
points made unaccessible to the particle motion have been removed. If
the cutout is not too severe, one will expect that the path integral
when computed in this new space leads to the same result as when
computed in the base space. In this case, the new space is called the
``{\it approximate space}'' of the problem. Because of the surgery
performed, its homotopic properties may be quite different from that
of the base space, and in particular may be non-trivial. Its distinct
topology is then referred to as the ``{\it approximate topology}'' of
the problem \cite{Sch2}.

\smallskip
When the above surgery results in the removal of a single point, which
is sufficient to transform a space with a trivial topology into a
multiply connected space, the value of the path integral is not
affected when moving to the approximate space. This point is generally
a singularity of some kind.  It may be where an exterior physical
constraint such as a localised field has been applied, or it may be
the singular point of a set of coordinates chosen to cover the base
space.

\smallskip
An example of a base space constraint by a localised field is
encountered in the Aharanov-Bohm effect \cite{KL,Sch2,AB}. In this
case a vector potential ${\mathbf A}$ is localised on a straight line
of $\R^3$. If this line is removed from the base space, the resulting
approximate space will be multiply connected. Its homotopic properties
enable one to write the propagator as an infinite sum of effective
propagators, where the sum is taken over the winding number with
respect to the above line.

\smallskip
The origin of the 2D plane covered by circular coordinates is the
simplest example of a singular coordinate point. If the origin is
removed from the plane, the topology of the corresponding approximate
space becomes that of $\R_+\times\s^1$, because the polar angle has an
intrinsic periodical structure. The propagator may then be decomposed
according to the winding number with respect to the origin, and
expressed in terms of free propagators in circular coordinates
\cite{KL,TV,Pol}.

\bigskip
The concept of approximate topology is not only useful when
topological or physical constraints are applied on a system, but also
when a particular calculation technique or method is considered.

\smallskip
Firstly, it is useful when WKB approximations are used or when
asymptotic solutions of a problem are considered. One may find that
either of these solutions break down and become singular at some
points (for example at the turning points of a potential), although
the exact solution is regular there. By removing these points, the
topology of the base space is modified and one may then resort to the
concept of approximate topology \cite{AA,BM}.

\smallskip
Secondly, if one considers a curved spacetime, one may prefer to work
in its Euclidean section for simplicity.  The topology of the
Euclidean section may be quite different from the one of the
Lorentzian section, as for example in the Schwarzschild black-hole
case \cite{LF}. Although it is widely believed that the value of a
path integral is not modified when moving to the Euclidean section,
this has never been proved in general to the best of my knowledge.

\smallskip
Thirdly, the geodesic structure may transform the topology of the base
space. For example, a star curves spacetime in such a way that two
points in its space projection $\R^3$ are connected by an infinite
number of geodesics. This give a multiply connected structure to the
spacetime although the set $\R^3$ is simply connected, even when the
star itself is removed from $\R^3$.

\bigskip
I have enumerated several problems for which the concepts of
approximate space and approximate topology are useful to consider when
calculating path integrals. However, I believe that for one of these
problems, these concepts are not very well-suited or appropriate. This
happens when the base space is covered by a set of coordinates
admitting one or several singular points.  In this case, I believe
that one should avoid adopting the point of view that it is the base
space which is transformed into an approximate space by removing the
coordinate singularities, and rather use the concept of the {\it
coordinates configuration space}, and work in this space instead. The
reason for doing this is that a choice of coordinates is a rather
passive one, in so far that it does not require a modification of the
base space in itself.

\smallskip
The coordinates configuration space is the entire set of values taken
by the chosen coordinates. In this space, there are no restrictions to
the particle motion stemming from the chosen set of coordinates, when
these coordinates cover the whole space. It contains also the
coordinate values of the singular points and so, contrary to the
approximate space, it is a closed space in general. Since a
singularity may correspond in general to a family of points in the
coordinates configuration space, the topology of this space may still
be quite different from that of the base space. For the same reason,
there may also be an ambiguity regarding the actual particle
trajectory in the coordinates configuration space when it moves
through a coordinate singularity in the base space.

\smallskip
As an example, we consider again the 2D plane covered by circular
coordinates.  Its origin is included in the configuration space, and
may be parametrised by the set of values $\{(r,\theta) \,\mid\,r=0
\text{ and } \theta\in[0,2\pi)\}$, where $r$ is the radius and
$\theta$ the polar angle. The configuration space in these coordinates
is an half-infinite cylinder whose radius is equal to unity, and where
its boundary, a circle, parametrises the origin. If the particle goes
through the origin in the base space, its trajectory on this circle is
not {\it a priori} determined.

\smallskip
From now on I shall reserve the terms ``{\it approximate space}'' and
``{\it approximate topology}'' for an ad hoc or {\it active} change of
space resulting from the choice of working in another space than the
base space, and use the terms ``{\it associated space}'' and ``{\it
associated topology}'' for the coordinates configuration space. In
this sense, as it will become clear in the present paper, a change of
coordinates (i.e.~a {\it passive} transformation) implies a change of
associated topology when this transformation possesses singular
points.

\begin{figure*}
\centerline{
\epsfysize=2.7truein
\epsfbox{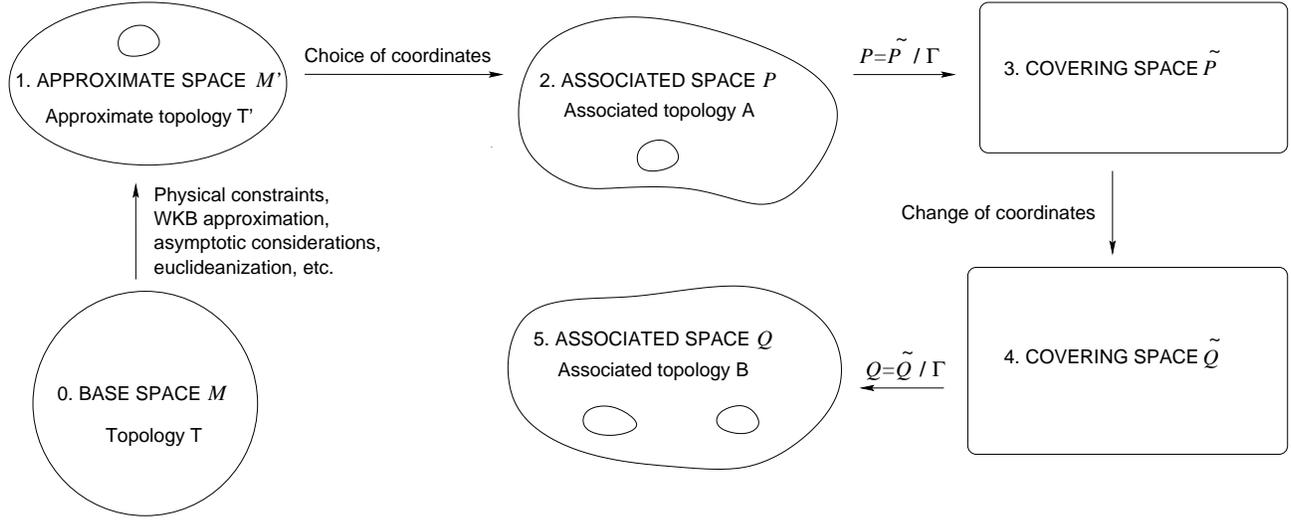}}
\mbox{\vspace{0mm}}
\caption{The relationship between the base, approximate, associated
and covering spaces ($\Gamma$ is the holonomy group). A change of
coordinates admitting a singularity modifies the spacetime associated
topology.}
\label{fig:trans}
\end{figure*}

\bigskip
The concepts of approximate and associated spaces are illustrated in
Fig.~\ref{fig:trans}, where it is shown also how they can be combined
to account for the various situations one may encounter. In a first
stage, one may choose to transform the base space into an approximate
space to take into account the physical constraints or in order to use
a given calculation technique. In a second stage, after a choice of
coordinates has been made to cover the approximate space, one
considers the coordinates associated space. Path integrals are
calculated in its universal covering space because it is simply
connected. This space is considered in the third stage. In the fourth
stage, a change of coordinates may then be performed, resulting in a
distinct covering space. In the last and fifth stage, one rebuilds the
associated space in the new coordinates and study its possibly new
topology.

\smallskip
In the literature, different approaches have been considered in
curved backgrounds. 

\smallskip
Troost and Van Dam \cite{TV}, in their study of Rindler spacetime,
considered its Euclidean section to compute the path integral and thus
worked in an Euclidean approximate space. They considered firstly
Euclidean Cartesian coordinates for which the associated topology is
trivial. The path winding number with respect to the bifurcation of
the event-horizons (i.e.~the origin of Euclidean spacetime) is then
introduced, and the propagator is expressed as a sum over this winding
number. One sums path integrals with a modified action as in
Ref.~\cite{Pol}, which amounts to working in circular coordinates
(i.e.~Euclidean Rindler coordinates in that case) for which the
associated topology is multiply connected.  These winding number
dependent path integrals are then calculated by solving the
corresponding Schr\"odinger equation. The thermal properties of the
propagator are then highlighted.

\smallskip
Hartle and Hawking \cite{HH} considered a black-hole background and
continued analytically the Kruskal time coordinates to obtain a
positive definite metric. The associated space has a non-trivial
topology because of the periodic structure of the imaginary Kruskal
time coordinate. They used the path integral to obtain the boundary
conditions satisfied by the propagator. From these boundary conditions
and from the Klein-Gordon equation, they deduced the analytical
properties obeyed by the propagator. In this way the thermal
properties of the propagator are revealed.

\bigskip
In an eternal black-hole background, one is generally interested in
computing the propagator in tortoise coordinates to capture the
thermal radiation. However, this task is made difficult by the fact
that these coordinates only cover the exterior region of the black
hole. The paths crossing the horizon cannot then be parametrised, and
as a consequence the path integral cannot be adequately expressed in
these coordinates. For this reason, one may choose to consider the
path integral in Euclidean Kruskal coordinates which cover the whole
black-hole spacetime, as did the above authors. However, in {\it
non-static} spacetimes such as in collapsing black-hole spacetimes
\cite{Sy}, the Euclidean approach is simply useless because in these
cases the periodical structure of the Kruskal type coordinates is
missing in the Euclidean section. A strict Lorentzian approach is then
required. Furthermore, the physical interpretation of the results in
the Euclidean section is not always obvious.

\smallskip
The philosophy of the present paper is to remain in the Lorentzian
section of the considered spacetimes throughout, and to attempt to
calculate the path integral in a direct way. This has been done in
{\it non-static} backgrounds in Ref.~\cite{OV}. The goal of this work
is to extend the previous analysis to {\it static} backgrounds. I
shall first consider the static Gui spacetime \cite{Gui1}, which I
believe is the simplest spacetime model incorporating event-horizons
which exhibits thermal properties. This vanishing curvature spacetime
is made up of four Minkowski quadrants glued together along the
``event-horizons''. The analysis shall then be extended to the more
realistic case of the Schwarzschild black hole. The similarities and
differences between the static Gui spacetime and the Schwarzschild
black-hole spacetime are discussed.

\smallskip
The key point of the Lorentzian approach of the present paper is the
concept of associated topology discussed above. As shown in
Ref.~\cite{OV}, this is different in coordinates related by a
non-analytical transformation. In this method, {\it one allows the
tortoise type coordinates to take complex values in order to cover the
entire spacetime}. The resulting complex associated space is multiply
connected essentially because these coordinates are singular at the
event-horizon, in contrast to the Kruskal associated space which is
homotopically trivial. A path crossing the event-horizon has a well
defined winding number with respect to the event-horizon in the
complex tortoise associated space. The propagator can then be written
as a sum of path integrals over this winding number which can be
evaluated in some instances. The thermal structure of the propagator
becomes then obvious. The thermal properties of the vacuum in a
black-hole background, for example, can thus be seen to follow from
topological considerations.

\bigskip
The second section is devoted to a review of the path integral
formalism and to the definition of the notations used in the present
paper. In the third section, this formalism is applied to the static
Gui spacetime. The path integral is computed there according to the
Lorentzian approach. The propagator between points belonging to
different or similar quadrants of this spacetime are obtained and the
thermal and hermitian properties of these propagators are analysed. In
the fourth section, the Schwarzschild black hole is investigated and
the results obtained are generalised. The propagator is computed
exactly far away from the black hole in the 2D case and some results
are given in the 4D case.

\section{Review of the path integral formalism}

A relativistic quantum particle is allowed to go virtually both
forwards and backwards in time \cite{HO}, contrary to the
non-relativistic case. In addition to the spacetime parameters, it is
then necessary to introduce a supplementary parameter $s$, the
so-called fifth parameter, to parametrise the path of a particle. This
may be interpreted in the massive case as the proper time of the
particle, and in general it plays the same role than that of the time
parameter in the non-relativistic framework. This fifth parameter
allows to describe more complex processes than in the non-relativistic
case, for example a spontaneous creation of particles.

\smallskip
A Schr\"{o}dinger equation for which $s$ is the evolution parameter
can be written for a spinless relativistic particle moving on a curved
spacetime. This one-particle dynamical equation does not depend
explicitly on the mass $m$ of the particle, and this enables one to
treat the massless and massive cases on an equal footing. It is
related to the Klein-Gordon equation by an imaginary Laplace transform
with respect to $s$, whose conjugate variable is $m^2$. The propagator
for this Schr\"{o}dinger equation can be written as a quantum
mechanical path integral as in the non-relativistic case. The
time-ordered two-point correlation function of the underlying quantum
field theory may then be obtained from this propagator.

\smallskip
As explained in Section \ref{sec:intro}, the path integral formalism
is quite useful in spaces endowed with a non-trivial topology. In this
context, the set over paths on which the sum is taken is of crucial
importance in the definition of the path integral. This set can
naturally be fixed by the space itself. Its choice is equivalent to
the one of the propagator boundary conditions when solving directly
the Schr\"{o}dinger equation, and to the choice of the vacuum state in
quantum field theory.

\bigskip
To fix the ideas, one considers a connected curved spacetime $\Mc$,
endowed with the metric $g$, which may be covered entirely with a set
of coordinates denoted by $q$. The associated space $\Qc$ in these
coordinates is defined to be the set of values taken by $q$, and the
induced metric on $\Qc$ is denoted by $g_q$. The space $\Qc$ should
always be connected and may be multiply connected in general.

\smallskip
The total probability amplitude for a quantum particle of mass $m$ to
move within $\Mc$ from an initial point $q_i\in\Qc$ to a final point
$q_f\in\Qc$ is given by the propagator $G_q(q_i;q_f;m^2)$ in $q$
coordinates. In curved spacetimes, the propagator for a spinless
relativistic particle satisfies to the Klein-Gordon equation
\begin{multline}
\left(\hbar^2\Box+\hbar^2\xi R + m^2\right)\,G_q\left(q';q'';m^2\right) 
\\[2mm]
= \ \parallel\!g\!\parallel^{-1/2}\delta(q'-q''),
\label{KGeq}
\end{multline}
where $\Box=\nabla^\mu\nabla_\mu$ if $\nabla_\mu$ denotes the
covariant derivative associated to the metric $g$, where $\delta$ is
the Dirac function, and where $\xi=0$ and $\xi=(D-2)/[4(D-1)]$
correspond to the minimal and conformal coupling respectively if $D$
is the spacetime dimension. One introduces a $s$-dependent propagator
$K_q(q';q'';s)$ by the imaginary Laplace transform
\begin{equation}
G_q \left(q';q'';m^2\right) = \frac{i}{\hbar}\int^\infty_0 ds\, 
\exp\left(-im^2s/\hbar\right)\,K_q \left(q';q'';s\right).
\label{defK}
\end{equation}
By postulating that the propagator $G_q(q_i;q_f;m^2)$ vanishes if
$m^2<0$, one obtains by inverting this last equation,
\begin{multline}
\theta\left(s\right) K_q\left(q';q'';s\right) \\[2mm]
= \frac{1}{2\pi i}\int^\infty_0 dm^2\, 
\exp\left(im^2s/\hbar\right)\,G_q\left(q';q'';m^2\right),
\label{defKinv}
\end{multline}
where $\theta$ is the step or Heaviside function. One then checks from
Eq.~(\ref{KGeq}) that $K_q$ satisfies to the Schr\"odinger equation
\begin{equation}
i\hbar\partial_s K_q\left(q';q'';s\right)=
\hbar^2\left(\Box +\xi R\right)\,K_q\left(q';q'';s\right),
\label{Schr}
\end{equation}
if and only if the $s$-parameter boundary condition for the propagator
is given by
\begin{equation}
\lim_{s\rightarrow0} K_q\left(q';q'';s\right)=\
\parallel\!g\!\parallel^{-1/2}\delta(q'-q'').
\label{BCs}
\end{equation}
The propagator $K_q(q_i;q_f;s)$ is interpreted as the probability
amplitude for the particle to move from the initial end point $q_i$ to
the final end point $q_f$ within a parameter time $s$. It is clear
that $K_q\left(q';q'';s\right)=K_q\left(q'';q';s\right)$.

\bigskip
One now introduces the one-particle Hilbert space ${\cal H}$ and the
localised states $\mid\!q\!>\ \in {\cal H}$, where $q\in\Qc$, which
satisfy to the orthogonality relation
\begin{equation}
<q'\mid q''> \ =\ \parallel\!g\!\parallel^{-1/2}\delta(q'-q'').
\label{OG}
\end{equation}
The operator of evolution $U(s)$ acting on ${\cal H}$ is defined from
the propagator $K_q$ by
\begin{align}
K_q\left(q';q'';s\right) = \ <q''\mid U(s)\mid q'>.
\label{defU}
\end{align}
It satisfies by definition to $U(s)^{-1}=U(-s)$. Since $U(0)=I$,
Eqs.~(\ref{OG}) and (\ref{defU}) imply consistently the $s$-boundary
condition, Eq.~(\ref{BCs}).  The evolution operator is unitary,
i.e.~$U(s)^\dagger=U(s)^{-1}$, if and only if
\begin{equation}
K_q\left(q';q'';s\right)^*=K_q\left(q'';q';-s\right).
\label{KHermiticity}
\end{equation} 
Furthermore, from Eq.~(\ref{defK}), we see that this last equation is
satisfied if and only if\footnote{The Hamiltonian $H$ is defined by
$U(s)=\exp\left(-isH/\hbar\right)$. From Eq.~(\ref{defK}), one obtains
$G=\left(H+m^2\right)^{-1}$. The operator $H$ is hermitian if and only
if $U(s)$ is unitary, and $G$ is hermitian if and only $H$ is
hermitian (the issues regarding the domain of the operators are not
considered here).}
\begin{equation}
G_q\left(q';q'';m^2\right)^*=G_q\left(q'';q';m^2\right).
\end{equation} 
In other words, $K_q$ describes a unitary evolution in the fifth
parameter if and only if $G_q$ is a hermitian operator.

\bigskip
The general form of $K_q$ is obtained by solving directly the
Schr\"{o}dinger equation, Eq.~(\ref{Schr}), and is given in
Ref.~\cite{BP} by
\begin{multline}
K^{vac}_q\left(q';q'';s\right) = \frac{1}{(4\pi\hbar is)^{D/2}}\,
\sqrt{\Delta(q';q'')}\ F(q';q'';s) 
\\  \quad \times \sum_{\gamma\in\Qc}
\exp\left[\frac{i}{\hbar}\frac{\sigma_g(q';q'';\gamma)^2}{4s}\right],\,
\label{GFK}
\end{multline}
where $\gamma$ denotes an arbitrary geodesic joining the end points
$q'$ and $q''$, $\sigma_g(q';q'';\gamma)$ is the proper arc length
between $q'$ and $q''$ along the particular geodesic $\gamma$,
$\Delta(q';q'')$ is the Van Vleck-Morette determinant and
$F(q';q'';s)$ is a function whose general expression is unknown. The
definition of the biscalars $\Delta(q';q'')$ and $F(q';q'';s)$ can be
found in Ref.~\cite{BP}. For our purpose, it is sufficient to realise
that these two functions are equal to unity in a flat spacetime
covered by any set of curvilinear coordinates. One notices that the
explicit dependence of Eq.~(\ref{GFK}) on the spacetime dimension $D$
is rather trivial.

\smallskip
A distinguished geodesic $\gamma_0$ in $\Qc$ is the one whose length
$\sigma(q';q'';\gamma_0)$ tends to zero when the end point $q'$
approaches $q''$. Its contribution has necessarily to be included in
the sum in Eq.~(\ref{GFK}) in order that the $s$-boundary condition,
Eq.~(\ref{BCs}), is satisfied. It is the only contribution which is
singular when the parameter $s$ vanishes. The contributions of all the
others geodesics individually satisfy to the Schr\"{o}dinger equation,
Eq.~(\ref{Schr}), and do not modify the $s$-boundary condition. In
consequence, one can {\it a priori} include in the sum in
Eq.~(\ref{GFK}) only those geodesics one wishes besides the geodesic
$\gamma_0$. The spacetime boundary conditions will be determined
crucially by this choice. However, instead of having to estimate
whether the resulting spacetime boundary conditions are physically
acceptable or not, one rather choses as a rule to include in
Eq.~(\ref{GFK}) the contributions of {\it all} the geodesics {\it
contained within $\Qc$}. This simple rule is legitimised by the fact
that a particle has, by definition, {\it virtual} access to all the
points of the {\it connected} associated space $\Qc$.

\smallskip
The particular solution in Eq.~(\ref{GFK}) defines a vacuum state
$\mid\!vac\!>$ of the underlying quantum field theory. This vacuum
defines a Fock space ${\cal F}$ which may be related to the associated
space $\Qc$. Notice that Fock spaces corresponding to distinct
associated spaces may not be necessarily unitarily equivalent
\cite{Um}.

\smallskip
The simplest example of a propagator is the one in a flat spacetime
covered by Minkowski coordinates. In this case, the geodesic joining
the two end points is unique. The relevant vacuum is the Minkowski one
denoted by $\mid\! M\!>$ and one has from Eq.~(\ref{GFK})
\begin{equation}
\begin{split}
K^M_q\left(q';q'';s\right) &\equiv K_0\left(q';q'';s\right) \\
&= \frac{1}{(4\pi\hbar is)^{D/2}}\,
\exp\left[\frac{i}{\hbar}\frac{(q''-q')^2}{4s}\right],
\label{K0}
\end{split}
\end{equation}
where $K_0$ denotes the free propagator.

\bigskip
In the general case, the propagator $K_q(q_i;q_f;s)$ is written in a
rather symbolic way as a sum over the paths $[q]$ contained within
$\Qc$ and joining the end points $q_i$ and $q_f$ within a parameter time
$s=s_f-s_i$,  
\begin{equation}
K^{vac}_q\left(q_i;q_f;s\right) =
\sideset{}{_{g_q}}\sum_{\substack{q_i\rightarrow q_f \\ [ q]\,\in\,\Qc}}
\exp\left( \frac{i}{\hbar} S_{g_q}[q]\right),
\label{SPq}
\end{equation}
where the covariant action $S_{g_q}[q]$ is given by
\begin{equation}
S_{g_q}[q] = \frac{1}{4}\int_{s_i}^{s_f}\,d\omega\,
g_{\mu\nu}(q)\,\dot{q}^\mu\,\dot{q}^\nu.
\end{equation}
It is conjectured that the vacuum state $\mid\!vac\!>$ defined in this
sum is the same as the one defined by Eq.~(\ref{GFK}).

\bigskip
If the space $\Qc$ is multiply connected, one introduces its covering
space $\Qct$ \cite{LL} which is always simply connected. To define it,
one considers the {\it homotopic classes} of paths of $\Qc$. By
definition, the paths of a given homotopic class can be deformed
continuously into one another, but it is not possible to do so for
paths belonging to distinct homotopic classes. The {\it holonomy
group} $\Gamma$ is the set of all the homotopic classes. By
definition, the covering space $\Qct$ contains the points denoted by
$\qt^\nu$, where $q$ is an arbitrary point of $\Qc$ and where the
index $\nu$ ranges over all the elements of the holonomy group
$\Gamma$. These points are called the {\it images} in $\Qct$ of the
point $q\in\Qc$. One has then $\Qc=\Qct/\Gamma$, and the elements of
$\Gamma$ can thus also be thought of as applications relating the
different image points $\qt^\nu$. For example, one defines the element
$\gamma^\nu\in\Gamma$ and the {\it base point} $\qt\equiv\qt^{\nu=0}$
in such a way that $\qt^\nu=\gamma^\nu(\qt)$. Paths in $\Qc$ with
identical initial and final end points but with distinct end points in
$\Qct$ belong to different homotopic classes. The covering space
$\Qct$ is endowed by the metric $\gt_\qt$ defined naturally by
$\gt(\qt^\nu)=g(q)$.

\smallskip
The sum over paths in Eq.~(\ref{SPq}) is rewritten in $\Qct$ by taking
into account its possible multiply connected topology, i.e.~by summing
over the classes of paths, or equivalently over the images
$\qt_f^\nu$,
\begin{equation}
\sideset{}{_{g_q}}\sum_{\substack{q_i\rightarrow q_f \\ [q]\,\in\,\Qc}}
\exp\left(\frac{i}{\hbar} S_{g_q}[q] \right) =
\sum_{\nu}\ \sideset{}{_{\gt_\qt}}
\sum_{\substack{\qt_i\rightarrow\qt_f^\nu \\ [\qt]\,\in\,\Qct}} 
\exp\left(\frac{i}{\hbar} S_{\gt_\qt}[\qt] \right).
\label{Pathint1}
\end{equation}
The propagator $\Kt_\qt$ in the covering space $\Qct$ is defined by
the sum over paths appearing in the right hand side (RHS) of this last
equation,
\begin{equation}
\Kt_\qt\left(\qt';\qt'';s\right)= \sideset{}{_{\gt_\qt}}
\sum_{\substack{\,\qt'\rightarrow \qt'' \\ [\qt]\,\in\,\Qct}}
\exp\left(\frac{i}{\hbar} S_{\gt_\qt}[\qt] \right).
\label{Pathint2}
\end{equation}
A new vacuum $\mid\!vac'\!>$ and a new propagator $K^{vac'}_q$ are
defined by the identification $\Kt_\qt\equiv K^{vac'}_q$. One thus
obtains the general and important result \cite{Ma,Do}
\begin{equation}
K^{vac}_q(q_i;q_f;s)=\sum_{\nu}K^{vac'}_q(\qt_i;\qt_f^\nu;s).
\label{sumnu}
\end{equation}

\smallskip
In general, the covering space $\Qct$ is not necessarily real but may
be complex. From Eq.~(\ref{Pathint1}), we deduce that if $\Qct$ is the
complex conjugate of itself, i.e.~$\Qct^*=\Qct$, evolution will be
unitary, i.e.~Eq.~(\ref{KHermiticity}) will be satisfied. There will be
end points for which the evolution is not unitary if and only if
$\Qct^*\not=\Qct$.

\bigskip
The sum over paths is defined as a path integral, i.e.~as an infinite
dimensional integral. Following Ref.~\cite{BP} one writes
\begin{equation}
\sum_{\substack{\qt'\rightarrow \qt'' \\ [\qt]\,\in\,\Qct}} 
\exp\left(\frac{i}{\hbar} S_{\gt_\qt}[\qt] \right) =
\underset{[ \qt]\,\in\,\Qct}{\int_{\qt'}^{\qt''}}
D_{\gt_\qt}[\qt] \, 
\exp\left(\frac{i}{\hbar} \widetilde{S}_{\gt_\qt}[\qt]\,\right),
\label{Pathint}
\end{equation}
where $\widetilde{S}_{\gt_\qt}[\qt]=
\int_{s_i}^{s_f}d\omega\,\widetilde{L}_{\gt_\qt}[\qt,\dot{\qt}]$,
and where the RHS of this last equation is defined by
\begin{multline}
\underset{[\qt]\,\in\,\Qct}{\int_{\qt'}^{\qt''}} \!
D_{\gt_\qt}[\qt] \,\, \exp\left(\frac{i}{\hbar}
\widetilde{S}_{\gt_\qt}[\qt]\right) 
= \lim_{N\rightarrow\infty} \!\!\!
\left\lgroup\frac{N}{4\pi\hbar is}\right\rgroup^{\!\!\!\frac{DN}{2}}\times 
\\[2mm] \prod_{j=1}^{N-1}
\int_{\Qct} d^D\qt_j \parallel\!\gt_\qt(\qt_j)\!\parallel^{\frac{1}{2}} 
\exp\left(\frac{i}{\hbar}\sum_{k=1}^N\int_{s_{k-1}}^{s_k} \!
d\omega\,\widetilde{L}_{\gt_\qt}[\,\qt,\dot{\qt}\,]\right),
\label{defPI} 
\end{multline}
where $s_j= s_i+js/N$, $\qt_j=\qt(s_j)$ ($j=0,1,...,N$), $N\in\N$. It
is assumed that each integral in the exponential is evaluated along
the image-geodesic connecting $\qt_{j-1}$ and $\qt_j$ which ensures
that the path integral is defined in a covariant way.  This rule
implies that the Lagrangian
$\widetilde{L}_{\gt_\qt}[\qt,\dot{\qt}]$ is given by
\begin{equation}
\widetilde{L}_{\gt_\qt}[\,\qt,\dot{\qt}\,] = \frac{1}{4}\,
\gt_{\mu\nu}(\qt)\,\dot{\qt}^\mu\,\dot{\qt}^\nu -\hbar^2(\xi-1/3)\,R(\qt).
\end{equation}
With the definition of Eq.~(\ref{defPI}), the term $\hbar^2 R/3$ must be
added to the usual Lagrangian to take into account the effect of the
curvature in the path integral in order to get to right value for the
propagator.

\section{Gui's spacetime}

\subsection{Spacetime model}

The $\eta$-$\xi$ spacetime of Gui, which I shall call Gui's spacetime,
was introduced in Ref.~\cite{Gui1}. Although this spacetime is rather
pathological in its nature, its study is instructive for many
reasons. Firstly, the path integral can be calculated in this
background in a direct way as shown in the present paper.  Secondly, I
believe that it is the {\it generic} example of a spacetime exhibiting
thermal properties in the sense that fields and states on this
spacetime behave as if they were immersed in a thermal bath contained
in a Minkowski background \cite{Gui1,Gui2}. And finally, this
spacetime shares some similarities with the Schwarzschild black-hole
spacetime, although their global causal properties are
different. These points are clarified in this section and in the next
one.

\begin{figure*}
\centerline{
\epsfysize=2.5truein
\epsfbox{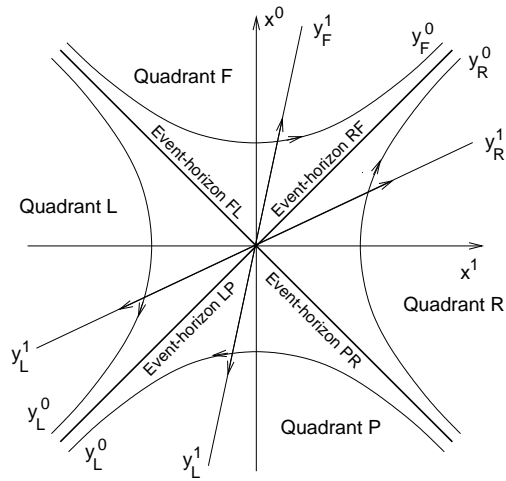}}
\mbox{\vspace{0mm}}
\caption{The Kruskal associated space $\Xc$ of Gui's spacetime. The
Kruskal coordinates $x$ and the tortoise coordinates $y_a$ are shown
($a=L,F,R,P$). The arrows indicate the direction of increasing
values.}
\label{fig:gui}
\end{figure*}

\smallskip
The Gui spacetime is a vanishing scalar curvature spacetime with a
non-trivial structure. In $D$ dimensions, it is defined by the line
element
\begin{equation}
ds^2_h = \frac{1}{\kappa^2}\,\frac{dx^+\,dx^-}{x^+\,x^-} - (d\xf)^2,
\label{Gds}
\end{equation}
where $\kappa>0$, $x^\pm=x^0\pm x^1$, $\xf\in\R^n$ ($D=n+2$), and
where $h$ denotes the Gui metric. The $x$ coordinates are Kruskal type
coordinates. This metric is rather pathological since it is singular
on the two hyperplanes given by $x^+=0$ and by $x^-=0$. These shall be
called ``event-horizons'' for this reason. They divide Gui's spacetime
into four quadrants, which are individually isomorphic to the
Minkowski spacetime (see below). These are denoted by $R,F,L$ and $P$
(the right, future, left and past quadrants); see
Fig.~\ref{fig:gui}. Each of these quadrants is causally disconnected
from the others in the {\it classical} sense, because the proper
distance of an event in one of these quadrants to the bordering
``event-horizons'' is infinite.  It is thus not possible for an
observer located in a given quadrant to infer the existence of the
others quadrants by performing a classical experiment. However, a {\it
quantum} experiment in a given quadrant may {\it a priori} be
influenced by the presence of the others, because in a quantum
framework a particle may virtually tunnel through the
``event-horizons''. One thus expects that the quadrants are causally
connected in this quantum sense only.

\subsection{Real tortoise coordinates}

In Gui's spacetime, the Kruskal associated space shall be denoted by
$\Xc$. It is given by
\begin{align}
\Xc &= \{\,(x^+,x^-,\xf)\in\R^D\,\},
\label{Xc}
\end{align}
and it is simply connected. Its covering space $\Xct$ is thus
isomorphic to it. In the covering space $\Xct$, the line element in
Eq.~(\ref{Gds}) is written in the form
\begin{equation}
ds^2_\hti = 
\frac{1}{\kappa^2}\,\frac{d\xt^+\,d\xt^-}{\xt^+\,\xt^-} - (d\xft)^2.
\label{Gdst}
\end{equation}

\smallskip
We now perform a change of coordinates in the covering space $\Xct$. In
each quadrant, one introduces the tortoise type coordinates $\yt_a$ by
\begin{align}
\text{in quad.~$R$:} & \quad  
\begin{cases}
\xt^0 = +(1/\kappa)\,\exp\left(\kappa \yt^1_\QR\right)\,
\sinh\left(\kappa \yt^0_\QR\right), \\[2mm]
\xt^1 = +(1/\kappa)\,\exp\left(\kappa \yt^1_\QR\right)\,
\cosh\left(\kappa \yt^0_\QR\right),
\end{cases}
\label{x(y)R} 
\end{align}
\vspace{-4mm}
\begin{align}
\text{in quad.~$F$:} & \quad
\begin{cases}
\xt^0 = +(1/\kappa)\,\exp\left(\kappa \yt^1_\QF\right)\,
\cosh\left(\kappa \yt^0_\QF\right), \\ [2mm]
\xt^1 = +(1/\kappa)\,\exp\left(\kappa \yt^1_\QF\right)\,
\sinh\left(\kappa \yt^0_\QF\right),
\end{cases} 
\label{x(y)F}
\end{align}
\vspace{-4mm}
\begin{align}
\text{in quad.~$L$:} & \quad
\begin{cases}
\xt^0 = -(1/\kappa)\,\exp\left(\kappa \yt^1_\QL\right)\,
\sinh\left(\kappa \yt^0_\QL\right), \\[2mm]
\xt^1 = -(1/\kappa)\,\exp\left(\kappa \yt^1_\QL\right)\,
\cosh\left(\kappa \yt^0_\QL\right),
\end{cases}
\label{x(y)L} \\[4mm]
\text{in quad.~$P$:} & \quad
\begin{cases}
\xt^0 = -(1/\kappa)\,\exp\left(\kappa \yt^1_\QP\right)\,
\cosh\left(\kappa \yt^0_\QP\right), \\ [2mm]
\xt^1 = -(1/\kappa)\,\exp\left(\kappa \yt^1_\QP\right)\,
\sinh\left(\kappa \yt^0_\QP\right),
\end{cases}
\label{x(y)P}
\end{align}
and by $\xft=\yft_a$ ($a=R,F,L,P$)\footnote{These transformations can
also be written in the form:
\begin{align}
\text{in quad.~$R$:}&\quad
\begin{cases}
\xt^+ = +(1/\kappa)\,\exp\left(+\kappa \yt^+_\QR\right), \\[2mm]
\xt^- = -(1/\kappa)\,\exp\left(-\kappa \yt^-_\QR\right),
\end{cases}
\nonumber 
\\[4mm]
\text{in quad.~$F$:}&\quad
\begin{cases}
\xt^+ = +(1/\kappa)\,\exp\left(+\kappa \yt^+_\QF\right), \\[2mm]
\xt^- = +(1/\kappa)\,\exp\left(-\kappa \yt^-_\QF\right),
\end{cases}
\nonumber 
\\[4mm]
\text{in quad.~$L$:}&\quad
\begin{cases}
\xt^+ = -(1/\kappa)\,\exp\left(+\kappa \yt^+_\QL\right), \\[2mm]
\xt^- = +(1/\kappa)\,\exp\left(-\kappa \yt^-_\QL\right),
\end{cases}
\nonumber 
\\[4mm]
\text{in quad.~$P$:}&\quad
\begin{cases}
\xt^+ = -(1/\kappa)\,\exp\left(+\kappa \yt^+_\QP\right), \\[2mm]
\xt^- = -(1/\kappa)\,\exp\left(-\kappa \yt^-_\QP\right).
\end{cases}
\nonumber
\end{align} }.
These transformations shall be called ``Rindler transformations''.  In
these equations, one has chosen the signs of $\yt^0_a$ and $\yt^1_a$
in such a way that the coordinates $\yt_a$ behave in a continuous way
when the corresponding spacetime event is moved from one quadrant to
another and when subjected to a slight displacement. For example, when
the event $(\xt^0,\xt^1)$ is moved towards the origin
$(\xt^0,\xt^1)=(0,0)$, one has $\yt^1_a\rightarrow-\infty$ in any
quadrant $a$. In a similar way, the coordinates $\yt^0_R$ and
$\yt^0_F$, for example, are both positive in any small region
containing a segment of the ``event-horizon'' $RF$. The tortoise
coordinates of a point on the ``event-horizon'' $RF$ or $LP$ are
$(\yt_a^0,\yt_a^1)=(+\infty,-\infty)$ when $a=R,F$ or $a=L,P$
respectively, and the ones of a point on the ``event-horizon'' $PR$ or
and $FL$ are $(\yt^0_a,\yt^1_a)=(-\infty,-\infty)$ when $a=R,P$ or
$a=L,F$ respectively. The couple $(\yt^0_a,\yt^1_a)=(c,-\infty)$
($c\in\R$, $a=R,F,L,P$) parametrises necessarily the bifurcation or
origin $(\xt^0,\xt^1)=(0,0)$.

\smallskip
In tortoise coordinates, the line element is the Minkowski one in any
quadrant,
\begin{align}
ds^2_\hti = d\yt^+_a\,d\yt^-_a - (d\yft_a)^2, \qquad a=R,F,L,P.
\end{align}
This shows that each of the quadrants is isomorphic to Minkowski
spacetime as stated above. The tortoise associated spaces $\Yc_a$ are
given by
\begin{align}
\Yc_a &= \{\,(y^+,y^-,\yf)\in\R^D\,\}, \qquad a=R,F,L,P.
\end{align}
They are simply connected and thus isomorphic to their covering spaces
$\Yct_a$ ($a=R,F,L,P$). The full tortoise covering space is given by
$\Yct_R\cup\Yct_F\cup\Yct_L\cup\Yct_P$, but it is not connected. This
implies that a path crossing an ``event-horizon'' cannot be
parametrised with only one set of tortoise coordinates, and that a
path integral cannot be expressed in these coordinates. In this sense,
the parametrisation of Gui's spacetime in terms of {\it real} tortoise
coordinates is not satisfactory.

\subsection{Complex tortoise coordinates}

As shown in this section, it is possible to parametrise the entire Gui
spacetime with only one set of tortoise coordinates, denoted by $\yt$,
if one allows them to take complex values. These coordinates take
their values in a connected complex covering space, denoted by $\Yct$.
One requires that in the {\it entire} Gui spacetime they satisfy
exclusively to the transformation given in Eq.~(\ref{x(y)R}),
\begin{equation}
\begin{cases}
\xt^0 = (1/\kappa)\,\exp\left(\kappa\yt^1\right)\,
\sinh\left(\kappa\yt^0\right), \\[2mm]
\xt^1 = (1/\kappa)\,\exp\left(\kappa\yt^1\right)\,
\cosh\left(\kappa\yt^0\right),
\end{cases}
\label{x(y)c}
\end{equation}
$\forall\,(\xt^0,\xt^1)\in\R^2$, $\forall\,\yt\in\Yct$.

\smallskip
To construct the space $\Yct$, one first considers the reciprocal of
this last transformation given in quadrant $R$ by
\begin{align}
\begin{cases}
\yt^0 &= \displaystyle\frac{1}{2\kappa}
\ln\left(\displaystyle-\frac{\xt^+}{\xt^-}\right), \\[3mm]
\yt^1 &= \displaystyle\frac{1}{2\kappa}\ln\left(-\kappa^2\xt^+\xt^-\right).
\end{cases}
\label{recip}
\end{align}
The functions $\yt^0=\yt^0(\xt^0,\xt^1)$ and
$\yt^1=\yt^1(\xt^0,\xt^1)$ are then continued analytically from
quadrant $R$ to the other quadrants by performing complex rotations of
$180^\circ$ in the $\xt^\pm$ complex planes to connect the positive
and negative values of $\xt^\pm$. Since these functions depend on {\it
two} complex variables, their analytical continuations will not
necessarily be unique.  In the covering space, it is natural to treat
the logarithm as a multivalued function. A cut is thus not fixed in
the $\xt^\pm$ complex planes, i.e.~the argument of the complex
variables is not bounded and takes any value ranging from $-\infty$ to
$+\infty$. The logarithm function is consequently defined here by
\begin{equation}
\ln\left(\xt^\pm\right) = 
\ln\left\vert\xt^\pm\right\vert + i\arg\left(\xt^\pm\right),
\end{equation}
where $\arg\left(\xt^\pm\right)\in\R$.  As a point of departure, we
now assume that in quadrant $R$ the values of $\yt^0$ and $\yt^1$ are
given by Eq.~(\ref{recip}) where ${\rm Im}\,\yt^0={\rm Im}\,\yt^1=0$.

\smallskip
One decides first whether the analytic continuation is performed to
the other quadrants according to the sequence $R\rightarrow
F\rightarrow L \rightarrow P$ or to the opposite one $R\rightarrow
P\rightarrow L \rightarrow F$ (in short $(R,F,L,P)$ and $(R,P,L,F)$
respectively). The results do not actually depend on the chosen
sequence.  According to the first sequence, one continues in the first
step in the $\xt^+$ complex variable from quadrant $R$ to quadrant
$F$. According to the second sequence, the analytic continuation is
performed in the first step with respect to the $\xt^-$ complex
variable from quadrant $R$ to quadrant $P$.

\smallskip
To fix the ideas, we choose the first sequence. When continuing
analytically from quadrant $R$ to quadrant $F$, the sign of $\xt^-$ is
reversed and becomes positive. They are two ways of implementing this
change of sign in the $\xt^-$ complex plane: either we add or subtract
$\pi$ to the argument of $\xt^-$. The values of the logarithm are
different in these two cases,
\begin{equation}
\ln\left(-\xt^-\right) = \ln\left(\xt^-\right) \pm i\pi.
\end{equation}
The analytic continuation can thus be performed either in the
anticlockwise or clockwise directions of the $\xt^-$ complex
plane. There are two alternatives. The choice of one alternative
determines how the extension is performed from quadrant $R$ to
quadrant $F$, and from quadrant $L$ to quadrant $P$ as well.
 
\smallskip
Next we analytically continue from quadrant $F$ to quadrant $L$. The
sign of $\xt^+$ is reversed in this process and becomes negative. The
analytic continuation is done this time in the $\xt^+$ complex
plane. Again there are two ways of implementing this change of sign:
either we continue in the anticlockwise or clockwise directions of the
$\xt^+$ complex plane. One has
\begin{equation}
\ln\left(\xt^+\right) = \ln\left(-\xt^+\right) \pm i\pi.
\end{equation}
This choice determines how the continuation is performed from quadrant $F$
to quadrant $L$, and from quadrant $P$ back to quadrant $R$ as
well. Again, there are two alternatives. 

\smallskip
Thus there are in total four ways of analytically continuing the
functions $\yt^0(\xt^0,\xt^1)$ and $\yt^1(\xt^0,\xt^1)$ from quadrant
$R$ to the other quadrants. The different analytic continuations are
given in Table \ref{fig:param}. They are distinguished by the
directions in the $\xt^\pm$ complex planes with respect to which the
analytical continuations have been performed (the symbols $+$ and $-$
mean ``anticlockwise'' and ``clockwise'' respectively). The couple
$(+,-)$ for instance designates the analytical continuation which has
been performed in the anticlockwise direction of the $\xt^+$ complex
plane and in the clockwise direction of the $\xt^-$ complex plane. It
is not difficult to see that when the sequence $(R,F,L,P)$ is changed
to the sequence $(R,P,L,F)$, the directions of the analytical
continuation in both $\xt^\pm$ complex planes are reversed. For
example, the analytical continuation $(+,-)$ for the sequence
$(R,F,L,P)$ is identical to the analytical continuation $(-,+)$ for
the sequence $(R,P,L,F)$.

\smallskip
When the analytic continuation has been performed from quadrant $R$ to
the other quadrants according either to the sequence $(R,F,L,P)$ or
$(R,P,L,F)$, we arrive back in quadrant $R$. The values taken there by
either the function $\yt^0(\xt^0,\xt^1)$ or $\yt^1(\xt^0,\xt^1)$
obtained from the analytic continuation procedure may or may not be
different from their departure values. If these values are different,
the function will be a multivalued function; if they are not, it will
be a single valued function. In the former case, the analytic
continuation procedure is repeated an infinite number of times to
obtain all the relevant values. In Table \ref{fig:param}, we see that
either the function $\yt^0(\xt^0,\xt^1)$ or $\yt^1(\xt^0,\xt^1)$ is a
multivalued function, not both of them. Furthermore, the values of the
multivalued function differ by $i\beta\nu$, where $\beta=2\pi/\kappa$
and $\nu\in\Z$. From now on, the analytical continuations will always
be denoted with respect to the sequence $(R,F,L,P)$.

\bigskip
We now analyse in detail the analytical continuation denoted by
$(+,-)$. In that case, we see from Table \ref{fig:param} that the
complex coordinates $\yt^\nu\in\Yct$ can be identified with the real
coordinates $\yt_a\in\Yct_a$ in such a way that
\begin{align}
\begin{cases}
\left(\yt^0\right)^\nu = \yt^0 + i\beta\nu, \\[2mm]
\left(\yt^1\right)^\nu = \yt^1, 
\end{cases}
\label{ynu0}
\end{align}
where $\nu\in\Z$ and
\begin{align}
\text{in quad.~$R$:} \quad& 
\begin{cases}
\yt^0=\yt^0_\QR, \\[2mm]
\yt^1=\yt^1_\QR, \\[2mm]
\end{cases}
\label{baseR+-}\\[2mm]
\text{in quad.~$F$:} \quad&
\begin{cases}
\yt^0=\yt^0_\QF+i\beta/4, \\[2mm]
\yt^1=\yt^1_\QF-i\beta/4, \\[2mm]
\end{cases}
\label{baseF+-} 
\end{align}
\vspace{-4mm}
\begin{align}
\text{in quad.~$L$:} \quad&
\begin{cases}
\yt^0=\yt^0_\QL+i\beta/2, \\[2mm]
\yt^1=\yt^1_\QL, \\[2mm]
\end{cases}
\label{baseL+-} \\[2mm]
\text{in quad.~$P$:} \quad& 
\begin{cases}
\yt^0=\yt^0_\QP+i3\beta/4, \\[2mm]
\yt^1=\yt^1_\QP-i\beta/4, \\[2mm]
\end{cases}
\label{baseP+-}
\end{align}
where $\beta=2\pi/\kappa$ and $\yft^\nu=\yft=\yft_a$
($a=R,F,L,P$). Indeed, Eq.~(\ref{x(y)c}) and Eqs.~(\ref{baseR+-}) to
(\ref{baseP+-}) imply the Rindler transformations given in
Eqs.~(\ref{x(y)R}) to (\ref{x(y)P}).

\begin{figure*}
\centerline{
\epsfysize=2.5truein
\epsfbox{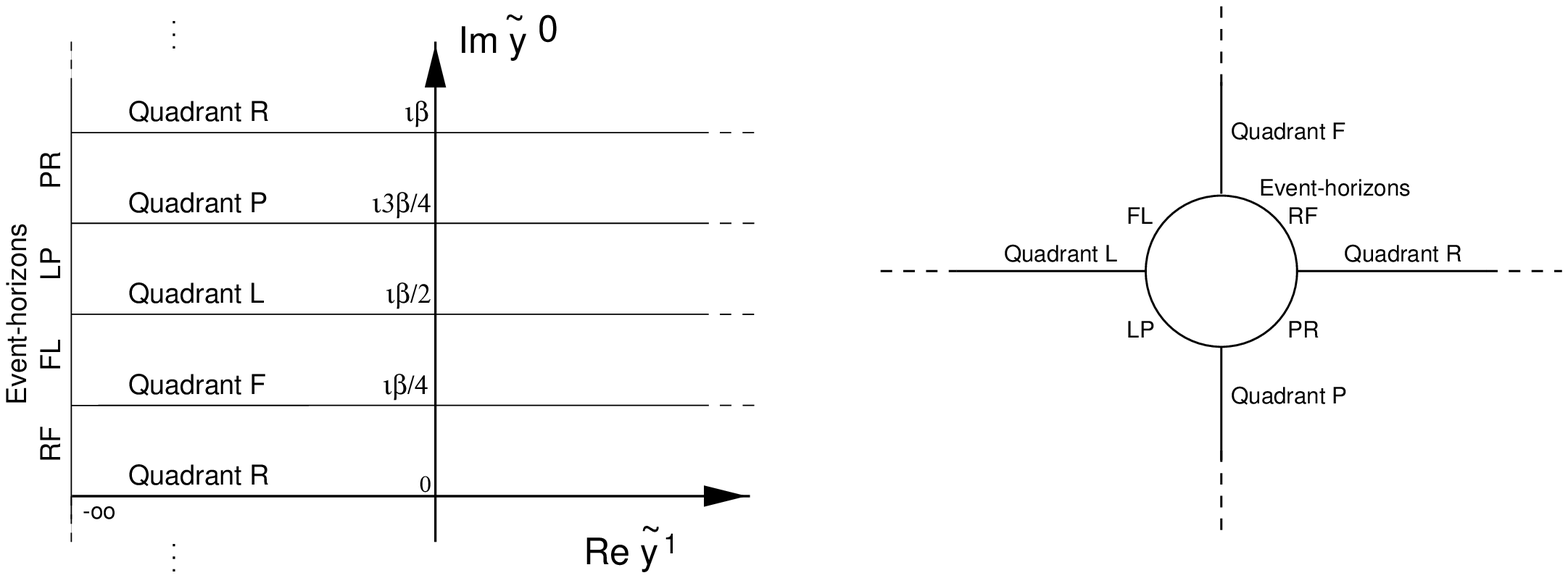}}
\mbox{\vspace{0mm}}
\caption{On the left hand side is shown a section of the complex
tortoise {\it covering} space $\Yct$ of Gui's spacetime.  The similar
section of the complex tortoise {\it associated} space $\Yc$ is shown
on the right hand side.}
\label{fig:covgui}
\end{figure*}

\smallskip
For this particular parametrisation, the covering space $\Yct$ should
therefore contain the points
\begin{equation}
(\yt^0,\yt^1,\yft)\in A_{\nu/4} \times B_\nu \times\R^n,
\label{setnu}
\end{equation}
where $\nu\in\Z$, and where the sets $A_\nu$ and $B_\nu$ are defined
by
\begin{align}
A_\nu &= \R + i\beta\nu, 
\label{defA}\\[2mm]
B_\nu &=
\left\{\begin{array}{ll}
\R, & \text{when $\nu$ is even}, \\[2mm]
B_-\equiv\R-i\beta/4, \qquad& \text{when $\nu$ is odd}.
\end{array}\right. 
\label{defB}
\end{align}
However, the covering space $\Yct$ cannot only be composed of the
joining of the sets in Eq.~(\ref{setnu}) over $\nu\in\Z$, because it
has to be connected. The regions which have not yet been parametrised
are the ``event-horizons'', which connect the four quadrants
together. They have to be included in the space $\Yct$ to make it
connected. One can convince oneself that the points of $\Yct$
parametrising the ``event-horizons'' are
\begin{equation}
(\yt^0,\yt^1,\yft)\in H_0\times H_1 \times \R^n,
\end{equation}
where
\begin{align}
H_0 &= \C\cup\{-\infty+i\R\}\cup\{+\infty+i\R\}, 
\label{defH1}\\[2mm]
H_1 &= -\infty+i[-\beta/4,0].
\label{defH2}
\end{align}
The set $\{+\infty+i\R\}\times H_1 \times \R^n$ parametrises the
``event-horizons'' $RF$ and $LP$ excluding the bifurcation. So do the
set $\{-\infty+i\R\}\times H_1 \times \R^n$ the ``event-horizons''
$FL$ and $PR$. The bifurcation is exclusively parametrised by the set
$\C\times H_1 \times \R^n$.  The full connected covering space $\Yct$
in complex tortoise coordinates is then given by
\begin{multline}
\Yct = \left\{\,(\yt^0,\yt^1,\yft)\in 
\rule{0mm}{6mm} \right. \\[2mm] \left.
\left[\bigcup_{\nu\,\in\,\Z} A_{\nu/4} 
\times B_\nu \times\R^n\right] 
\bigcup \left[H_0\times H_1 \times \R^n\right]\,\right\}.
\label{Yct}
\end{multline}
In summary, the sets $\bigcup_{\nu\,\in\,\Z} A_\nu \times\R\times\R^n$
and $\bigcup_{\nu\,\in\,\Z} A_{\nu+1/2}\times\R\times\R^n$ cover the
right and left quadrants respectively, and the sets
$\bigcup_{\nu\,\in\,\Z} A_{\nu+1/4}\times B_- \times\R^n$ and
$\bigcup_{\nu\,\in\,\Z} A_{\nu+3/4}\times B_- \times\R^n$ the future
and past quadrants. The set $H_0\times H_1\times\R^n$ parametrises the
``event-horizons'' including the bifurcation. A projection of the
complex covering space $\Yct$ is shown on the LHS of
Fig.~\ref{fig:covgui}.

\bigskip
For the sake of completeness, we now give the parametrisations
corresponding to the other analytic continuations of Table
\ref{fig:param}. The $(-,+)$ parametrisation is given by
Eq.~(\ref{ynu0}) where
\begin{align}
\text{in quad.~$R$:} \quad& 
\begin{cases}
\yt^0=\yt^0_\QR, \\[2mm]
\yt^1=\yt^1_\QR, \\[2mm]
\end{cases}
\label{baseR-+}\\[2mm]
\text{in quad.~$F$:} \quad&
\begin{cases}
\yt^0=\yt^0_\QF+i3\beta/4, \\[2mm]
\yt^1=\yt^1_\QF+i\beta/4, \\[2mm]
\end{cases}
\label{baseF-+} \\[2mm]
\text{in quad.~$L$:} \quad&
\begin{cases}
\yt^0=\yt^0_\QL+i\beta/2, \\[2mm]
\yt^1=\yt^1_\QL, \\[2mm]
\end{cases}
\label{baseL-+} \\[2mm]
\text{in quad.~$P$:} \quad& 
\begin{cases}
\yt^0=\yt^0_\QP+i\beta/4, \\[2mm]
\yt^1=\yt^1_\QP+i\beta/4, \\[2mm]
\end{cases}
\label{baseP-+}
\end{align}
where $\beta=2\pi/\kappa$ and $\yft^\nu=\yft=\yft_a$ ($a=R,F,L,P$). 

\smallskip
The $(+,+)$ parametrisation is given by
\begin{align}
\begin{cases}
\left(\yt^0\right)^\nu = \yt^0, \\[2mm]
\left(\yt^1\right)^\nu = \yt^1+ i\beta\nu, 
\end{cases}
\label{ynu1}
\end{align}
where $\nu\in\Z$ and
\begin{align}
\text{in quad.~$R$:} \quad& 
\begin{cases}
\yt^0=\yt^0_\QR, \\[2mm]
\yt^1=\yt^1_\QR, \\[2mm]
\end{cases}
\label{baseR++}\\[2mm]
\text{in quad.~$F$:} \quad&
\begin{cases}
\yt^0=\yt^0_\QF-i\beta/4, \\[2mm]
\yt^1=\yt^1_\QF+i\beta/4, \\[2mm]
\end{cases}
\label{baseF++} \\[2mm]
\text{in quad.~$L$:} \quad&
\begin{cases}
\yt^0=\yt^0_\QL, \\[2mm]
\yt^1=\yt^1_\QL+i\beta/2, \\[2mm]
\end{cases}
\label{baseL++} \\[2mm]
\text{in quad.~$P$:} \quad& 
\begin{cases}
\yt^0=\yt^0_\QP-i\beta/4, \\[2mm]
\yt^1=\yt^1_\QP+i3\beta/4; \\[2mm]
\end{cases}
\label{baseP++}
\intertext{while the $(-,-)$ parametrisation is given by
Eq.~(\ref{ynu1}) where}
\text{in quad.~$R$:} \quad& 
\begin{cases}
\yt^0=\yt^0_\QR, \\[2mm]
\yt^1=\yt^1_\QR, \\[2mm]
\end{cases}
\label{baseR--}\\[2mm]
\text{in quad.~$F$:} \quad&
\begin{cases}
\yt^0=\yt^0_\QF+i\beta/4, \\[2mm]
\yt^1=\yt^1_\QF+i3\beta/4, \\[2mm]
\end{cases}
\label{baseF--} 
\end{align}
\vspace{-4mm}
\begin{align}
\text{in quad.~$L$:} \quad&
\begin{cases}
\yt^0=\yt^0_\QL, \\[2mm]
\yt^1=\yt^1_\QL+i\beta/2, \\[2mm]
\end{cases}
\label{baseL--} \\[2mm]
\text{in quad.~$P$:} \quad& 
\begin{cases}
\yt^0=\yt^0_\QP+i\beta/4, \\[2mm]
\yt^1=\yt^1_\QP+i\beta/4, \\[2mm]
\end{cases}
\label{baseP--}
\end{align}
where $\beta=2\pi/\kappa$ and $\yft^\nu=\yft=\yft_a$ ($a=R,F,L,P$).
These parametrisations imply again with Eq.~(\ref{x(y)c}) the Rindler
transformations given in Eqs.~(\ref{x(y)R}) to (\ref{x(y)P}).

\subsection{Topology of $\Yc$ and winding number}

We now return to the parametrisation $(+,-)$ given in
Eqs.~(\ref{ynu0}) to (\ref{baseP+-}). The holonomy group $\Gamma$
of its corresponding covering space $\Yct$, Eq.~(\ref{Yct}), is given
by
\begin{align}
\Gamma = \{\,\gamma^\nu: 
(\yt^0,\yt^1,\yft) \mapsto (\yt^0+i\beta\nu,\yt^1,\yft),\nu\in\Z\,\}.
\end{align}
Points in $\Yct$ that are related by an element of $\Gamma$ obviously
correspond to a unique spacetime event. However, points corresponding
to the same spacetime event are not necessarily related by an element
of $\Gamma$. For example, the points of $H_0\times H_1\times\R^n$ all
describe the same spacetime event (i.e.~a point on the horizon), but
are not all related to each other by elements of $\Gamma$. This is
because the isometries of $\C^2\times\R^n$ defined by
$(\yt^0,\yt^1,\yft) \mapsto (\yt^0+ic,\yt^1,\yft)$, where $c\in\R$,
for which $H_0\times H_1\times\R^n$ is an invariant set, are not
defined {\it globally} in $\Yct$, and thus cannot be included in the
holonomy group $\Gamma$. This simple mathematical fact has deep
physical consequences.

\smallskip
The associated space $\Yc$ is defined by the quotient
$\Yc=\Yct/\Gamma$, and one of its projections is represented on the
RHS of Fig.~\ref{fig:covgui} for the parametrisation
$(+,-)$. Obviously, it is not a simply connected space, precisely
because the points of $\Yct$ describing the ``event-horizons'' have
not been all identified together. The topology of the horizon in the
space $\Yc$ is clearly that of a circle. A winding number may be
defined with respect to it, i.e.~with respect to the
``event-horizons.'' This shall be denoted by $\nu$.
 
\begin{figure*}
\centerline{
\epsfysize=2.5truein
\epsfbox{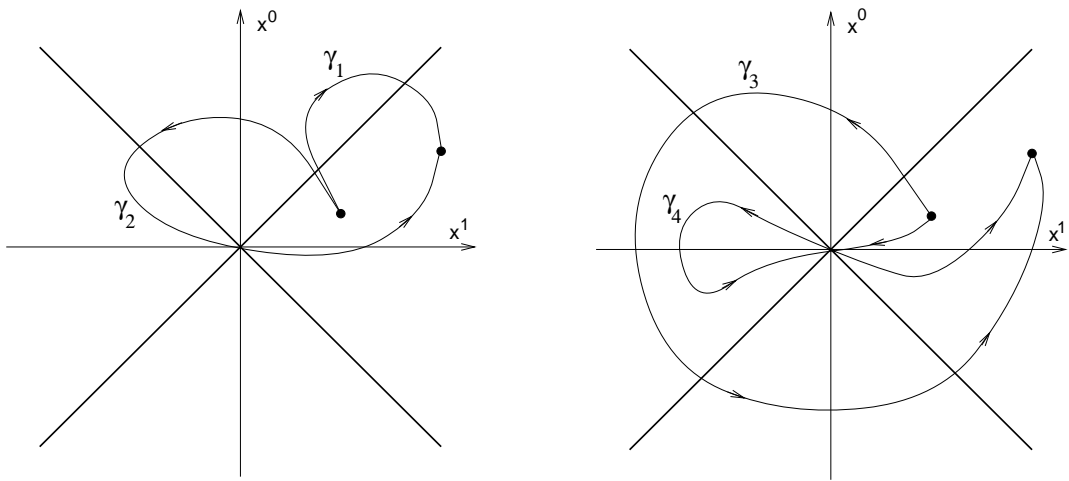}}
\caption{The paths $\gamma_1$ to $\gamma_4$ in the Kruskal associated
space $\Xct$.}  \mbox{\vspace{0mm}}
\label{fig:pathsx}
\end{figure*}

\begin{figure*}
\centerline{
\epsfysize=2.0truein
\epsfbox{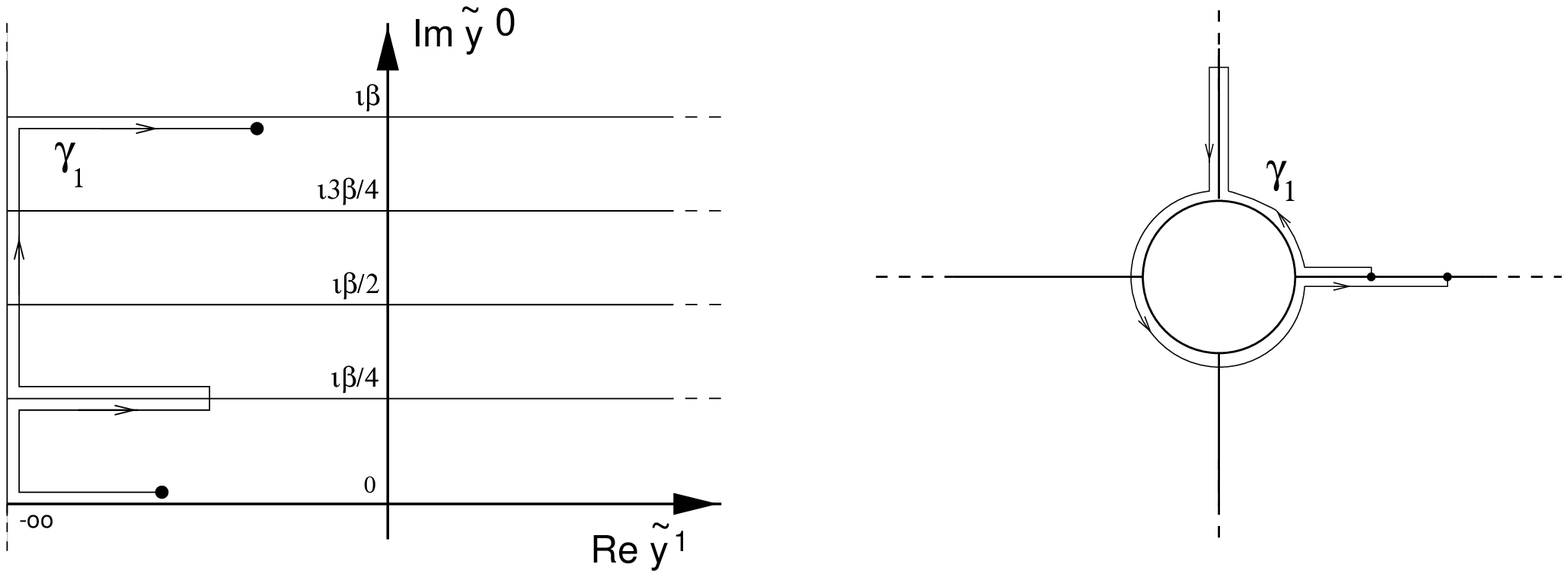}}
\mbox{\vspace{0mm}}
\centerline{
\epsfysize=2.0truein
\epsfbox{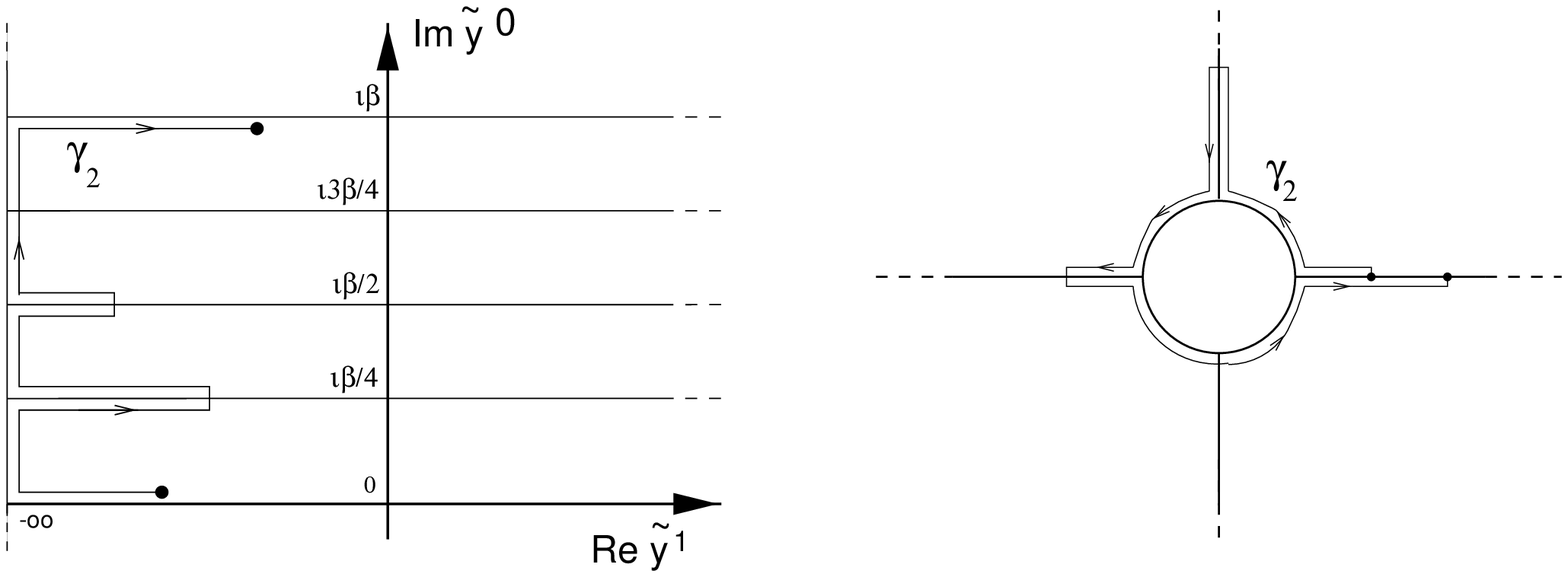}}
\mbox{\vspace{0mm}}
\caption{On the left hand side are shown the paths $\gamma_1$ and
$\gamma_2$ in the complex tortoise {\it covering} space $\Yct$. These
paths are shown in the complex tortoise {\it associated} space $\Yc$
on the right hand side.}
\label{fig:path12}
\end{figure*}

\begin{figure*}
\centerline{
\epsfysize=2.0truein
\epsfbox{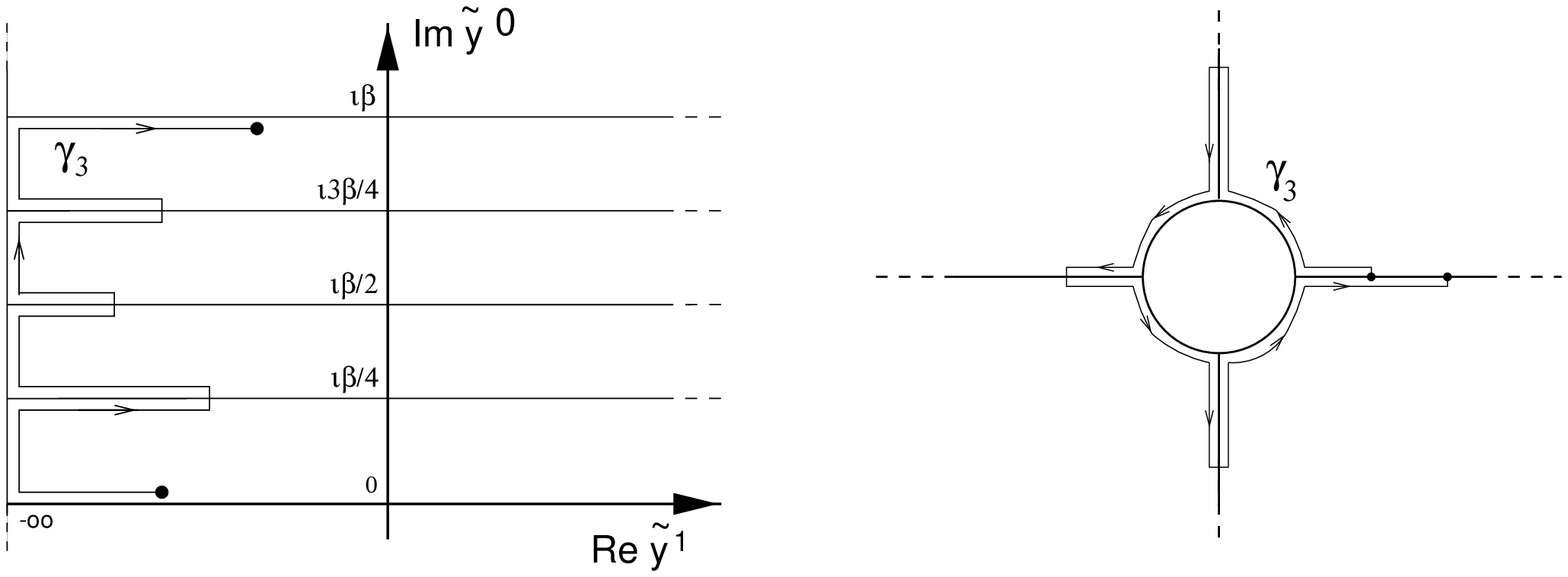}}
\mbox{\vspace{0mm}}
\centerline{
\epsfysize=2.0truein
\epsfbox{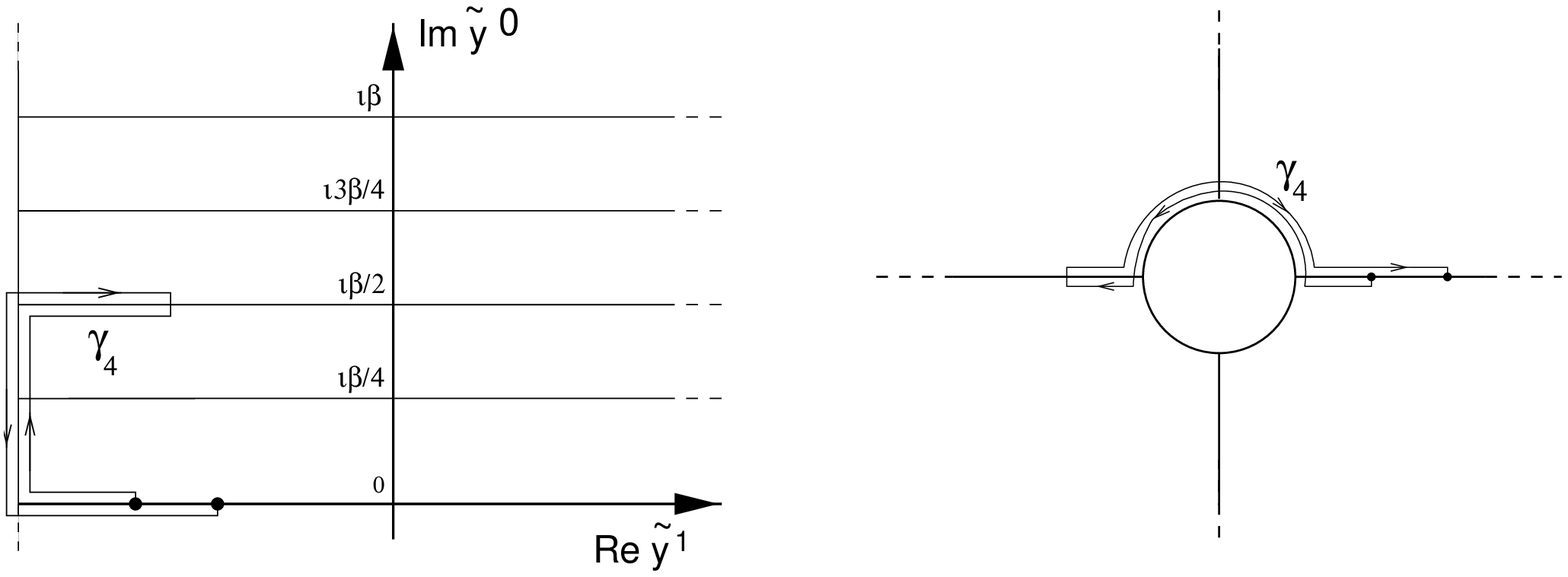}}
\mbox{\vspace{0mm}}
\caption{On the left hand side are shown the paths $\gamma_3$ and
$\gamma_4$ in the complex tortoise {\it covering} space $\Yct$. These
paths are shown in the complex tortoise {\it associated} space $\Yc$
on the right hand side.}
\label{fig:path34}
\end{figure*}

\bigskip
The winding number concept was used for the first time by Troost and
Van Dam \cite{TV} in the context of the eternal black-hole and Rindler
spacetimes to compute path integrals. The definition of the winding
number they introduced is easily extended to Gui's spacetime, in which
case it is defined in the {\it real} associated space $\Xc$ and with
respect to the origin $(x^0,x^1)=(0,0)$. It is distinct from the one I
have just introduced. To see clearly the differences between these and
to familiarise ourselves with them, I defined and represented four
paths, denoted by $\gamma_1$ to $\gamma_4$, in the Kruskal associated
space $\Xc$ (see Fig.~\ref{fig:pathsx}) and in the tortoise associated
and covering spaces $\Yc$ and $\Yct$ (see Figs.~\ref{fig:path12} and
\ref{fig:path34}). An advantage of the winding number defined with
respect to the ``event-horizons'' over the one defined with respect to
the origin is that the former is always well defined. In the Kruskal
associated space $\Xc$, we see that the path $\gamma_1$ crosses the
``event-horizon'' $RF$ twice and that its winding number with respect
to the origin vanishes. However, in the covering or associated spaces,
$\Yct$ or $\Yc$, we see that its winding number with respect to the
``event-horizons'' is +1. The path $\gamma_2$ crosses the
``event-horizons'' $RF$ and $FL$ once and then passes through the
origin before returning to the quadrant $R$. In this case, the winding
number with respect to the ``event-horizons'' is +1, and the one with
respect to the origin is not defined. The path $\gamma_3$ crosses the
four ``event-horizons'' $RF$, $FL$, $LP$ and $PR$ once. Its winding
numbers with respect to the ``event-horizons'' and to the origin are
both equal to +1. Finally, we see that the path $\gamma_4$ goes
through the origin twice. In this case, the winding number with
respect to the ``event-horizons'' vanishes, and the one with respect
to the origin is again not defined.

\bigskip
From Eq.~(\ref{Yct}), we see that the complex tortoise covering space
for the parametrisation $(+,-)$ is periodic in the imaginary time
direction. One draws the same conclusion for the parametrisation
$(-,+)$. The corresponding covering space can also be constructed in
the latter case and found to be similar but not identical to the
former case.  For the parametrisations $(+,+)$ and $(-,-)$, as one can
easily convince oneself, the complex tortoise covering space is
periodic in the imaginary {\it space} direction. The associated spaces
are similar but not identical in these two last cases as well.

\subsection{Applying the path integral formalism} \label{sec:pi}

From Eq.~(\ref{SPq}), the propagator in Gui's spacetime and in Kruskal
coordinates is expressed as a sum over paths in the associated space
$\Xc$ given in Eq.~(\ref{Xc}),
\begin{equation}
K^{Kr}_x\left(x_i,x_f;s\right) = \sideset{}{_{h_x}}
\sum_{\substack{x_i\rightarrow x_f \\ [ x]\,\in\,\Xc}}
\exp\left(\frac{i}{\hbar} S_{h_x}[ x]\right),
\label{SPx}
\end{equation}
where the metric $h_x$ in Kruskal coordinates is defined in
Eq.~(\ref{Gds}).  This sum over paths defines a Kruskal-like vacuum
$\mid\!Kr\!>$. It needs to be properly defined by using a principal
value because the metric and its determinant are singular at the
``event-horizons'' (see Appendix). 

\smallskip
In tortoise coordinates, we now consider the 16 propagators
$K^{Kr}_{y,ab}(y_{i,a};y_{f,b};s)$, where $a,b\in\{R,F,L,P\}$ and
where the end points $y_{i,a}$ and $y_{f,b}$ belong to the quadrants
$a$ and $b$ respectively. They describe the propagation from quadrant
$a$ to quadrant $b$ in tortoise coordinates. For simplicity, we shall
actually drop the subscript $a$ in $y_{i,a}$ when writting the
propagator since the quadrants are already specified as substricts of
the propagator. One writes thus $K^{Kr}_{y,ab}(y_{i,a};y_{f,b};s)
\equiv K^{Kr}_{ab}(y_i;y_f;s)$.  Since the propagator is a biscalar,
one has when $x_i$ and $x_f$ belong respectively to quadrants $a$ and
$b$,
\begin{align}
K^{Kr}_{ab}(y(x_i);y(x_f);s) = K^{Kr}_x \left(x_i;x_f;s\right)
\label{Kbiscalar}
\end{align}
where $y_i=y(x_i)$ and $y_f=y(x_f)$, if $y=y(x)$ is the inverse of the
Rindler transformations given in Eqs.~(\ref{x(y)R}) to
(\ref{x(y)P}). It is clear that $K^{Kr}_{ab}(y_i;y_f;s)=
K^{Kr}_{ba}(y_f;y_i;s)$.

\smallskip
To compute the sum over paths in Eq.~(\ref{SPx}), one performs a
change of coordinates to the complex tortoise coordinates covering the
entire spacetime. The relevant connected associated space to consider
is then $\Yc$ in any chosen parametrisation. From
Eq.~(\ref{Kbiscalar}), one has then
\begin{equation}
\sideset{}{_{h_x}}\sum_{\substack{x_i\rightarrow x_f \\ [x]\,\in\,\Xc}}
\exp\left(\frac{i}{\hbar} S_{h_x}[x]\right) =
\sideset{}{_\eta}\sum_{\substack{y_i\rightarrow y_f \\ [y]\,\in\,\Yc}}
\exp\left(\frac{i}{\hbar} S_\eta[y]\right),
\label{SPxy}
\end{equation}
where $\eta$ is the Minkowski metric.  When this sum over paths is
written in the complex tortoise covering space $\Yct$, one sums over
all the images of the final end point,
\begin{equation}
\sideset{}{_\eta}\sum_{\substack{y_i\rightarrow y_f \\ [y]\,\in\,\Yc}}
\exp\left(\frac{i}{\hbar} S_\eta[y]\right) = 
\sum_{\nu\in\Z} \underset{[\yt]\,\in\,\Yct}{\int_{\yt_i}^{\yt^\nu_f}} 
D_\etat[\yt]\,\exp\left(\frac{i}{\hbar}\widetilde{S}_\etat[\yt]\right),
\end{equation}
where $\etat=\eta$. Here $\nu$ is the winding number with respect to
the ``event-horizons''. This path integral is also badly defined and
needs to be redefined (see Appendix).  The contribution of the
homotopic class of winding number $\nu$ is the free propagator $\Kt_0$
with end points $\yt_i$ and $\yt_f^\nu$ (see Appendix),
\begin{equation}
\underset{[\yt]\,\in\,\Yct}{\int_{\yt_i}^{\yt^\nu_f}}
D_\etat[\yt]\,\exp\left(\frac{i}{\hbar}
\widetilde{S}_\etat[\yt]\right) = \Kt_0(\yt_i,\yt_f^\nu;s).
\label{Knu}
\end{equation}
One has $\Kt_0=K^M_y=K_0$, where $\mid\!M\!>$ is the Minkowski vacuum and
where $K_0$ is the free propagator; see Eq.~(\ref{K0}). By adding the
propagators of Eq.~(\ref{Knu}), the total propagator is obtained
\begin{equation}
K^{Kr}_{ab}\left(y_i;y_f;s\right)= \sum_{\nu\in\Z} K_0
\left(\yt_i;\yt^\nu_f;s\right).
\label{Gfinal}
\end{equation}

\bigskip
We now restrict ourselves to the parametrisation $(+,-)$, which is
periodic in the imaginary time direction; see Eq.~(\ref{ynu0}). In
this case, Eq.~(\ref{Gfinal}) becomes
\begin{equation}
K^{Kr}_{ab}\left(y_i;y_f; s\right)= \sum_{\nu\in\Z} 
K_0 \left(\yt_i;\yt^0_f+i\beta\nu,\yt_f^1,\yft_f; s\right).
\label{Gfinal*}
\end{equation}
Consequently, the propagators $K^{Kr}_{ab}$ are periodic in the
imaginary time coordinate,
\begin{equation}
K^{Kr}_{ab}\left(y_i;y^0_f,y^1_f,\yf_f; s\right) =
K^{Kr}_{ab}\left(y_i;y^0_f+i\beta\nu,y^1_f,\yf_f; s\right).
\end{equation}

\smallskip
The propagators $K^{Kr}_{ab}(y_i;y_f;s)$ are obtained by replacing
in Eq.~(\ref{Gfinal*}) the points $\yt_i$ and $\yt_f$ by their
base-point values given in Eqs.~(\ref{baseR+-}) to
(\ref{baseP+-}). One then deduces that
\begin{multline}
K^{Kr}_{_{RR}}\left(y_i;y_f;s\right) = 
K^{Kr}_{_{FF}}\left(y_i;y_f;s\right) 
\\[1mm] 
= K^{Kr}_{_{LL}}\left(y_i;y_f;s\right)
= K^{Kr}_{_{PP}}\left(y_i;y_f;s\right)
\\[1mm] 
=\sum_{\nu\in\Z} 
K_0 \left(y_i;y^0_f+i\beta\nu,y^1_f,\yf_f;s\right),
\label{KRR}
\end{multline}
\vspace{-5mm}
\begin{multline}
K^{Kr}_{_{RF}}\left(y_i;y_f;s\right) 
= K^{Kr}_{_{LP}}\left(y_i;y_f;s\right) \\[1mm] 
= \sum_{\nu\in\Z} K_0 \left( y_i;
y^0_f+i\beta(\nu+1/4),y^1_f-i\beta/4,\yf_f; s\right),
\label{KRF}
\end{multline}
\vspace{-5mm}
\begin{multline}
K^{Kr}_{_{RL}}\left(y_i;y_f;s\right) = 
K^{Kr}_{_{FP}}\left(y_i;y_f;s\right) 
\\[1mm]
= \sum_{\nu\in\Z} 
K_0 \left(y_i;y^0_f+i\beta(\nu+1/2),y^1_f,\yf_f; s\right),
\label{KRL}
\end{multline}
\vspace{-5mm}
\begin{multline}
K^{Kr}_{_{RP}}\left(y_i;y_f;s\right) =
K^{Kr}_{_{LF}}\left(y_i;y_f;s\right)
\\[1mm]
=\sum_{\nu\in\Z} 
K_0 \left(y_i;y^0_f+i\beta(\nu+3/4),y^1_f-i\beta/4,\yf_f; s\right),
\label{KRP}
\end{multline}
where the property $K_0(\yt_i+\zt;\yt_f;s)=K_0(\yt_i;\yt_f-\zt;s)$
($\zt\in\R^D$) has been used.  {\it The propagators in Eq.~(\ref{KRR})
describe a steady flux of thermal radiation of temperature
$T=\hbar\kappa/(2\pi k)$, where $k$ is the Boltzmann constant}.

\bigskip
The parametrisation $(-,+)$ is also periodic in the imaginary time
direction, so that Eq.~(\ref{Gfinal*}) is also true in this case. By
applying this equation to the base-point values given in
Eqs.~(\ref{baseR-+}) to (\ref{baseP-+}), one obtains furthermore the
decompositions,
\begin{multline}
K^{Kr}_{_{RF}}\left( y_i; y_f; s\right) 
\\[1mm]
=\sum_{\nu\in\Z} K_0 \left( y_i;
y^0_f+i\beta(\nu+3/4),y^1_f+i\beta/4,\yf_f; s\right),
\label{KRFbis}
\end{multline}
\vspace{-5mm}
\begin{multline}
K^{Kr}_{_{RP}}\left( y_i; y_f; s\right)
\\[1mm]
= \sum_{\nu\in\Z} K_0 \left( y_i;
y^0_f+i\beta(\nu+1/4),y^1_f+i\beta/4,\yf_f; s\right).
\label{KRPbis}
\end{multline} 
Since $K_0(\yt_i;\yt_f;s)=K_0(\yt_f;\yt_i;s)$ and
$K_0(\yt_i+\zt;\yt_f;s)=K_0(\yt_i;\yt_f-\zt;s)$, we deduce that the
decompositions in Eqs.~(\ref{KRFbis}) and (\ref{KRPbis}) are
equivalent to the ones in Eqs.~(\ref{KRF}) and (\ref{KRP})
respectively. From the results obtained in this section up to now, one
can check that the propagator is hermitian, i.e.~that
\begin{align}
K^{Kr}_{ab}\left(y_i;y_f;s\right)^* =
K^{Kr}_{ba}\left(y_f;y_i;-s\right),
\label{KHery}
\end{align}
$\forall\,a,b\in\{R,F,L,P\}$, as it should be since $\Yct^*=\Yct$ (see
discussion after Eq.~(\ref{sumnu})) .

\smallskip
The parametrisations $(+,+)$ and $(-,-)$ are periodic in the imaginary
space direction; see Eq.~(\ref{ynu1}). Consequently, we have from
Eq.~(\ref{Gfinal}),
\begin{equation}
K^{Kr}_y\left(y_i;y_f;s\right)= \sum_{\nu\in\Z} K_0
\left( \yt_i;\yt^0_f,\yt^1_f+i\beta\nu,\yft_f; s\right).
\label{Gfinalbis}
\end{equation}
In a similar way than above, and from the base-point values given in
Eqs.~(\ref{baseR++}) to (\ref{baseP++}) and (\ref{baseR--}) to
(\ref{baseP--}) respectively, we deduce from this last equation the
decompositions
\begin{align}
K^{Kr}_{_{RR}}\left( y_i; y_f; s\right) 
=\sum_{\nu\in\Z} 
K_0 \left( y_i; y^0_f,y^1_f+i\beta\nu,\yf_f; s\right),
\label{KRRter}
\end{align}
\vspace{-5mm}
\begin{multline}
K^{Kr}_{_{RF}}\left( y_i; y_f; s\right) 
\\[2mm]
= \sum_{\nu\in\Z} K_0 \left( y_i; 
y^0_f-i\beta/4, y^1_f+i\beta(\nu+1/4),\yf_f; s\right)
\label{KRFterA}
\end{multline}
\vspace{-5mm}
\begin{align}
= \sum_{\nu\in\Z} K_0 \left( y_i; 
y^0_f+i\beta/4,y^1_f+i\beta(\nu+3/4),\yf_f; s\right),
\label{KRFterB}
\end{align}
\vspace{-5mm}
\begin{multline}
K^{Kr}_{_{RL}} \left( y_i; y_f; s\right)
\\[1mm]
= \sum_{\nu\in\Z} 
K_0 \left( y_i; y^1_f, y^0_f+i\beta(\nu+1/2),\yf_f; s\right),
\label{KRLter}
\end{multline}
\vspace{-5mm}
\begin{multline}
K^{Kr}_{_{RP}}\left( y_i; y_f; s\right) 
\\[2mm]
= \sum_{\nu\in\Z} K_0 \left( y_i; 
y^0_f-i\beta/4, y^1_f+i\beta(\nu+3/4),\yf_f; s\right) 
\label{KRPterA}
\end{multline}
\vspace{-5mm}
\begin{align}
= \sum_{\nu\in\Z} K_0 \left(y_i; 
y^0_f+i\beta/4,y^1_f+i\beta(\nu+1/4),\yf_f; s\right).
\label{KRPterB}
\end{align}
The decompositions in Eqs.~(\ref{KRFterA}) and (\ref{KRFterB}) are
equivalent, so are those in Eqs.~(\ref{KRPterA}) and (\ref{KRPterB}).

\subsection{Non-static Gui's spacetime} \label{sec:UG}

We now review the non-static Gui spacetime and give some new
results. This spacetime has been introduced and studied in
Ref.~\cite{OV}. In $D$ dimension, it is defined by the metric
\begin{align}
ds^2_h = -\frac{dx^+\,dx^-}{\kappa\,x^-} -(d\xf)^2,
\label{GdsNS}
\end{align}
where $\kappa>0$, $x^\pm=x^0\pm x^1$, and $\xf\in\R^n$ ($D=n+2$). The
regions $I$ and $I\!I$ are defined by the half-planes $x^-<0$ and
$x^->0$ respectively. The ``event-horizon'' is located at $x^-=0$. Two
sets of tortoise type coordinates $y_\QI$ and $y_\QII$ are defined by
\begin{align}
&\text{in region $I$:} \quad&
x^-(y^-_\QI) & = - \exp\left(-\kappa y^-_\QI\right),
\label{x(y1)NS}  \\[2mm]
&\text{in region $II$:} \quad&
x^-(y^-_\QII) & = + \exp\left(-\kappa y^-_\QII\right),
\label{x(y2)NS}
\end{align}
and by $x^+=y^+_{\QI,\QII}$, $\xf=\yf_{\QI,\QII}$.  The complex
tortoise coordinates $\yt\in\Yct$ are defined in the covering space by
\begin{align}
\begin{cases}
\left(\yt^+\right)^\nu = \yt^+, \\[2mm] 
\left(\yt^-\right)^\nu = \yt^- +i\beta\nu,
\end{cases}
\label{NSynu}
\end{align}
where
\begin{align}
&\text{in region $I$:} &\quad&
\begin{cases}
\yt^+=y^+_\QI, \\[2mm]
\yt^-=y^-_\QI,
\end{cases} 
\label{NSbaseI}\\[2mm]
&\text{in region $I\!I$:} &\quad&
\begin{cases}
\yt^+=y^+_\QII, \\[2mm]
\yt^-=y^-_\QII+i\beta/2,
\end{cases}
\label{NSbaseII}
\end{align}
and by $\yft^\nu=\yft=\yf_{\QI,\QII}$. One deduces from
Eqs.~(\ref{x(y1)NS}) to (\ref{NSbaseII}) that
\begin{align}
\xt^-(\yt^-) =  - \exp\left(-\kappa\yt^-\right),
\label{x(y)cNS}
\end{align} 
in the joining of regions $I$ and $I\!I$. The complex covering space
$\Yct$ is then given by
\begin{align}
\Yct = \{\,(\yt^+,\yt^-,\yft)\in 
\R\times\bigcup_{\nu\,\in\,\Z} A_{\nu/2}\bigcup H \times \R^n\,\},
\label{ABNS}
\end{align}
where 
\begin{align}
A_\nu &=\R+i\beta\nu, \\[2mm]
H &=+\infty+i\R. 
\end{align}
The sets $\bigcup_{\nu\,\in\,\Z} A_{\nu}$ and $\bigcup_{\nu\,\in\,\Z}
A_{\nu+1/2}$ parametrise regions $I$ and $I\!I$ respectively. The axis
$H$ at infinity parametrises the ``event-horizon". 

\smallskip
Since the covering space $\Yct$ is the complex conjugate of itself,
quantum evolution is unitary in the fifth parameter, even through the
event-horizon. And because the covering space is periodic in the
imaginary $\yt^-$ direction, one has from Eq.~(\ref{sumnu})
\begin{align}
K^{Kr}_y\left(y_i;\,y_f;s\right) = \sum_{\nu\in\Z} 
K_0 \left(\yt_i;\yt^+_f,\yt^-_f+i\beta\nu,\yft_f;s\right),
\label{GfinalNS}
\end{align}
where $\beta=2\pi/\kappa$; see Ref.~\cite{OV}.  One then defines the 4
propagators $K^{Kr}_{ab}(y_i;y_f;s)$, where $a,b\in\{I,I\!I\}$ and
where $y_i$ and $y_f$ belong to regions $a$ and $b$ respectively. They
describe the propagation from region $a$ to region $b$. In a similar
way as in Sec.~\ref{sec:pi}, one considers the base-point values given
in Eqs.~(\ref{NSbaseI}) and (\ref{NSbaseII}).  From
Eq.~(\ref{GfinalNS}), the propagators are then given by
\begin{multline}
K^{Kr}_{_{I,I}} \left(y_i;y_f;s\right)
= K^{Kr}_{_{I\!I,I\!I}} \left(y_i;y_f;s\right)
\\[2mm] =\sum_{\nu\in\Z} 
K_0 \left(y_i;y^+_f,y^-_f+i\beta\nu,\yf_f;s\right),
\label{KII}
\end{multline}
\vspace{-5mm}
\begin{multline}
K^{Kr}_{_{I,I\!I}} \left(y_i;\,y_f;s\right)
\\[2mm] = \sum_{\nu\in\Z} 
K_0 \left(y_i;\,y^+_f,y^-_f+i\beta(\nu+1/2),\yf_f;s\right). 
\label{KIII}
\end{multline}

\section{Eternal black hole}

\subsection{Two-dimensional case}

\begin{figure*}
\centerline{\epsfysize=2.5truein\epsfbox{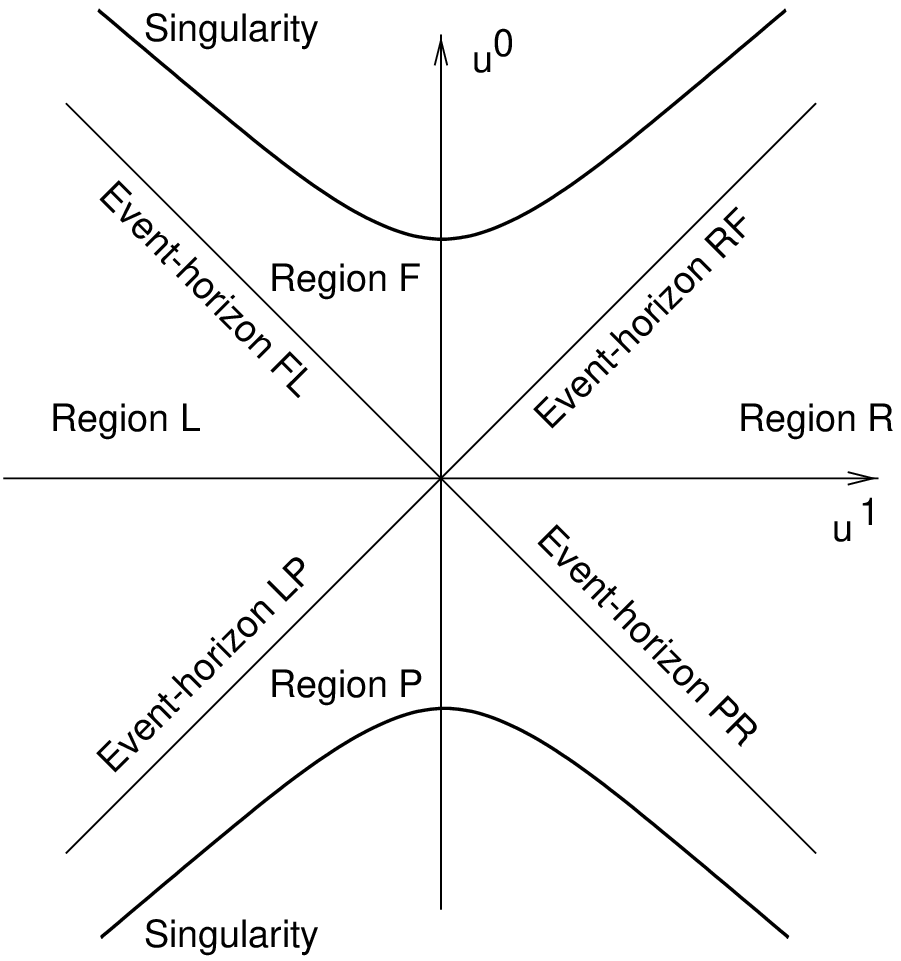}}
\mbox{\vspace{0mm}}
\caption{The Kruskal associated space $\Uc$ of the Schwarzschild
black-hole spacetime.  The regions $F$ and $P$ are the black-hole
interior regions, and the regions $R$ and $L$ are the black-hole
exterior regions.}
\label{fig:bh}
\end{figure*}

The 2D Schwarzschild black-hole line element is given by
\cite{BD}
\begin{equation}
ds^2_g = \frac{1}{\kappa^2}\,
\left[\,1-\frac{2M}{r}\,\right]\,\frac{du^+\,du^-}{u^+\,u^-},
\label{Sds2D}
\end{equation}
where $\kappa=(4M)^{-1}$ and
\begin{align}
\left\vert u^+/u^-\right\vert &= \exp(t/2M),
\label{u21}\\[1mm]
u^+u^-  &= -(r-2M)\exp(r/2M),
\label{u22} 
\end{align}
if $t$ and $r$ are the time and radius coordinates; see
Fig.~\ref{fig:bh}. The $u$ coordinates are Kruskal coordinates. The
tortoise coordinates $v_a$ ($a=R,F,L,P$) are defined by the Rindler
transformations, i.e.~by Eqs.~(\ref{x(y)R}) to (\ref{x(y)P}) when $x$
is replaced by $u$ and $y_a$ by $v_a$. The scalar curvature vanishes
asymptotically far away from the black hole. The Kruskal and tortoise
associated spaces, denoted by $\Uc$ and $\Vc_a$ respectively
($a=R,F,L,P$), are given by
\begin{align}
\Uc &=\{\,u\in\R^2\mid r(u)>0\,\}, && \\
\Vc_a &=\{\,v_a\in\R^2\,\}, &\quad a&=R,L, \\
\Vc_b &= \{\,v_b\in\R^2\mid r(v)>0 \,\}, &\quad b&=F,P.
\end{align}
These spaces have been restricted to the spacetime events for which
the radius is strictly positive. The spaces $\Vc_F$ and $\Vc_P$
parametrise the interior regions of the black hole. The spaces $\Uc$
and $\Vc_a$ are clearly isomorphic to their covering spaces $\Uct$ and
$\Vct_a$ ($a=R,F,L,P$).

\smallskip
The vacuum defined by the sum over paths,
\begin{equation}
K^{Kr}_u (u_i;u_f;s) = \sideset{}{_{g_u}} 
\sum_{\substack{u_i \rightarrow u_f \\ [ u]\in\,\Uc}} 
\exp\left(\frac{i}{\hbar} S_{g_u}[ u]\right),
\end{equation}
is the Kruskal vacuum $\mid\!Kr\!>$.

\begin{figure*}
\centerline{
\epsfysize=2.5truein
\epsfbox{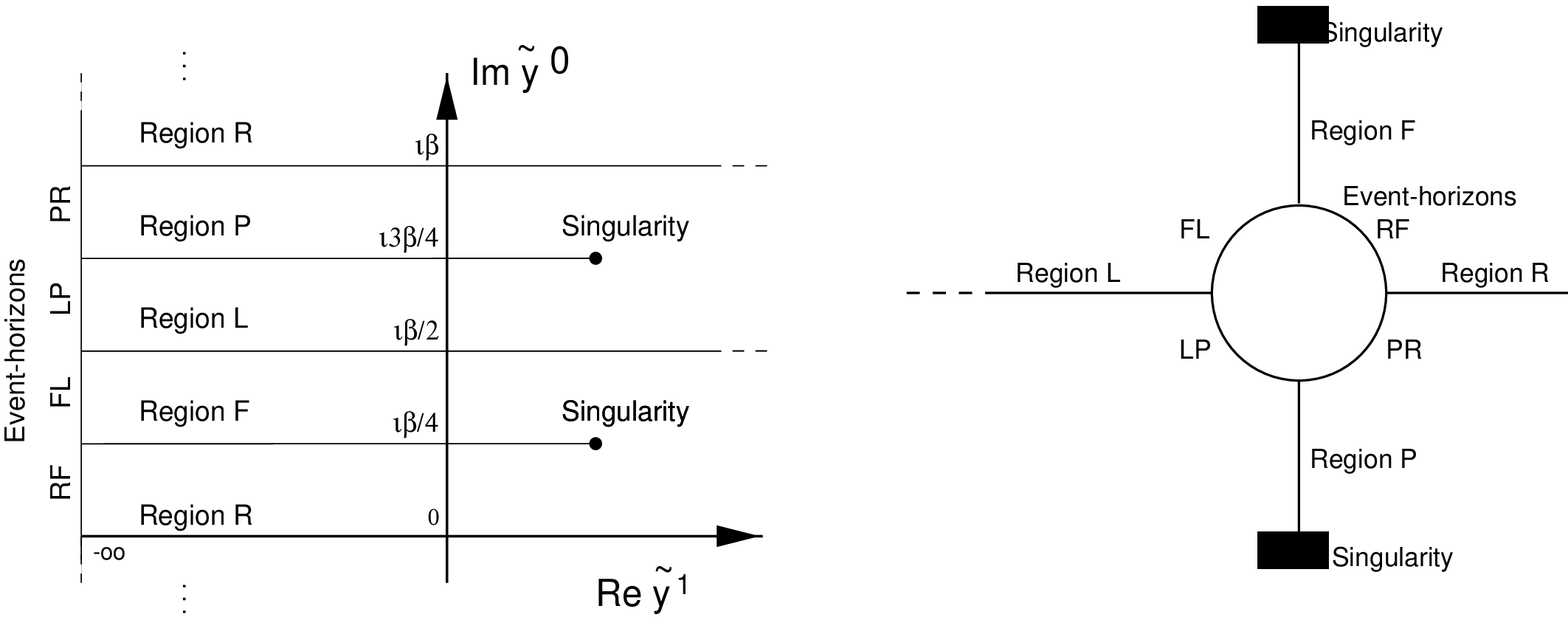}}
\mbox{\vspace{0mm}}
\caption{On the left hand side is shown a section of the complex
tortoise {\it covering} space $\Vct$ of the Schwarzschild black-hole
spacetime.  The similar section of the complex tortoise {\it
associated} space $\Vc$ is shown on the right hand side.}
\label{fig:covbh}
\end{figure*}

\bigskip
As in Gui's case, one introduces complex tortoise coordinates. For the
parametrisation $(+,-)$, these are defined by Eqs.~(\ref{ynu0}) to
(\ref{baseP+-}), where $\yt$ is replaced by $\vt$, and $\yt_a$ by
$\vt_a$. The Rindler transformation in the covering space then becomes
\begin{equation}
\begin{cases}
\ut^0 = (1/\kappa)\,\exp\left(\kappa\yt^1\right)\,
\sinh\left(\kappa\vt^0\right), \\[2mm] 
\ut^1 = (1/\kappa)\,\exp\left(\kappa\yt^1\right)\,
\cosh\left(\kappa\vt^0\right);
\end{cases}
\label{u(v)c}
\end{equation}
see Eq.~(\ref{x(y)c}).  The connected covering space $\Vct$ in complex
tortoise coordinates is consequently given by
\begin{multline}
\Vct = \Biggl\{\,(\vt^0,\vt^1)\in 
\Biggr.\\[2mm] \left. 
\left[\bigcup_{\nu\,\in\,\Z} A_{\nu/4} \times B_\nu\right] 
\bigcup \left[H_0\times H_1\right]\mid r(v)>0\, \right\},
\label{Vct2d}
\end{multline}
where $A_\nu$, $B_\nu$, $H_0$ and $H_1$ are defined in
Eqs.~(\ref{defA}), (\ref{defB}), (\ref{defH1}) and (\ref{defH2}); see
also Eq.~(\ref{Yct}). The space $\Vct$ is shown along with the
corresponding associated space $\Vc$ in Fig.~\ref{fig:covbh}.

\smallskip
Since the covering space is invariant under the action of the elements
of the holonomy group $\Gamma$, given by
\begin{align}
\Gamma = \{\,\gamma^\nu: 
(\vt^0,\vt^1) \mapsto (\vt^0+i\beta\nu,\vt^1),\nu\in\Z\,\},
\end{align}
the propagator satisfies
\begin{equation}
\Kt_v\left(\vt_i;\vt^0_f+i\beta\nu,\vt^1_f;s\right) 
= \Kt_v\left(\vt^0_i-i\beta\nu,\vt^1_i;\vt_f;s\right),
\label{K2d}
\end{equation}
where $\nu\in\Z$. One obtains furthermore from Eq.~(\ref{sumnu})
\begin{equation}
K^{Kr}_v\left(v_i;v_f;s\right) = \sum_{\nu\in\Z} K^B_v
\left(\vt_i;\vt^0_f+i\beta\nu,\vt^1_f;s\right),
\label{Sfinalbis}
\end{equation}
where $\mid\!B\!>$ is the Boulware vacuum \cite{BD}. Contrary to Gui's
case, the propagator $K^B_v$ is {\it not} the free one, since $g_v$ is
not the Minkowski metric.  From this last equation, one obtains
similar results than that of Gui's spacetime for $D=2$,
i.e.~Eqs.~(\ref{KRR}) to (\ref{KRP}) remain true if $K_0$ is replaced
by $K^B_v$. One has then
\begin{multline}
K^{Kr}_{_{RR}}\left( v_i; v_f; s\right)= \sum_{\nu\in\Z} 
K^B_v\left( v_i; v^0_f+i\beta\nu,v^1_f; s\right),
\label{KRR2d} 
\end{multline}
\vspace{-5mm}
\begin{multline}
K^{Kr}_{_{RF}}\left( v_i; v_f; s\right) \\[2mm]
= \sum_{\nu\in\Z} 
K^B_v \left( v_i;
v^0_f+i\beta(\nu+1/4),v^1_f-i\beta/4; s\right),
\label{KRF2d}
\end{multline}
\vspace{-5mm}
\begin{align}
K^{Kr}_{_{RL}}\left( v_i; v_f; s\right) = 
\sum_{\nu\in\Z} 
K^B_v \left( v_i;v^0_f+i\beta(\nu+1/2),v^1_f;s\right),
\label{KRL2d}
\end{align}
\vspace{-8mm}
\begin{multline}
K^{Kr}_{_{RP}}\left(v_i;v_f; s\right) \\[2mm]
=\sum_{\nu\in\Z} 
K^B_v \left( v_i;
v^0_f+i\beta(\nu+3/4),v^1_f-i\beta/4; s\right),
\label{KRP2d}
\end{multline}
and other results similar to those of Sec.~\ref{sec:pi}.

\bigskip
We now concentrate our attention on a connected region $\Rc$ located
far away from the black hole (i.e.~where $r\gg 2M$, $\forall\,t\in\R$). In
$\Rc$, the line element, Eq.~(\ref{Sds2D}), becomes
\begin{equation}
ds^2_g \approx \frac{1}{\kappa^2}\frac{du^+\,du^-}{u^+\,u^-}\cdot
\label{SdsR2D}
\end{equation}
Under the identifications $u\equiv x$, {\it the asymptotic form of the
2D black-hole line element in $\Rc$, Eq.~(\ref{SdsR2D}), is identical
to the Gui metric, Eq.~(\ref{Gds}), when $D=2$}.

\smallskip
In the 2D Gui spacetime, there is only one geodesic joining any two
points in the Kruskal associated space. If one assumes that these
points are located in $\Rc$ and that they are on the same side of the
black hole, this unique geodesic will be entirely contained within
$\Rc$.  Since spacetime is asymptotically flat in $\Rc$, one may write
from Eq.~(\ref{GFK}),
\begin{equation}
K^{Kr}_u\left(u_i;u_f;s\right) \approx \frac{1}{4\pi\hbar is}\,
\exp\left[\frac{i}{\hbar}\frac{\sigma_{g_u}(u_i,u_f)}{4 s}\right].
\label{ClAp2D}
\end{equation}
The exact equivalent expression can be written in Gui's spacetime,
\begin{equation}
K^{Kr}_x\left(x_i;x_f;s\right) = \frac{1}{4\pi\hbar is}\,
\exp\left[\frac{i}{\hbar}\frac{\sigma_{h_x}(x_i,x_f)}{4 s}\right].
\label{ClAp2DG}
\end{equation}

\smallskip
From now on, I shall use the notation $\approxeq$ to compare two
functions which tend to each other in a given limit when their
arguments are identified, and also the term ``{\it asymptotically
equivalent}'' to refer to this property.  For example, from
Eqs.~(\ref{Gds}) and (\ref{SdsR2D}), the metrics $h_x$ and $g_u$ are
asymptotically equivalent in $\Rc$, i.e.~$h_x\approxeq g_u$, when the
coordinates $x$ and $u$ are identified, i.e.~when $x\equiv u$.

\smallskip
Returning to Eqs.~(\ref{ClAp2D}) and (\ref{ClAp2DG}), and since under
our assumption the unique geodesic is entirely contained within $\Rc$,
$h_x\approxeq g_u$ implies that $\sigma_{h_x}\approxeq\sigma_{g_u}$ in
$\Rc$.  Consequently, the propagators in Eqs.~(\ref{ClAp2D}) and
(\ref{ClAp2DG}) are asymptotically equivalent in this region as well,
i.e.~$K^{Kr}_x(x_i;x_f;s) \approxeq K^{Kr}_u(u_i;u_f;s)$ in $\Rc$. In
terms of sum over paths, we have
\begin{equation}
\sideset{}{_{g_u}}\sum_{\substack{u_i \rightarrow u_f \\ [
u]\in\,\Uc}} \exp\left(\frac{i}{\hbar} S_{g_u}[
u]\right) \approxeq \sideset{}{_{h_x}}\sum_{\substack{x_i
\rightarrow x_f \\ [ x]\,\in\,\Xc}} \exp\left(\frac{i}{\hbar}
S_{h_x}[ x]\right).
\label{SPgh2D}
\end{equation}
Thus although one sums over paths that leave the region $\Rc$ and
cross the event-horizon, the propagators of the two problems are
nevertheless asymptotically equivalent. As it will become clear below,
this is not true in higher dimensional spacetimes for topological
reasons. Equations (\ref{KRR}) and (\ref{SPgh2D}) implies then that
the propagator $K^{Kr}_v$ in tortoise coordinates is asymptotically
equal in region $\Rc$ to the propagator of a steady flux of thermal
radiation. In terms of Schwarzschild coordinates, one has
\begin{equation}
K^{Kr}_v\left(t_i,r_i;t_f,r_f;s\right) \approx \sum_{\nu\in\Z} 
K_0 \left(t_i,r_i;t_f+i\beta\nu,r_f;s\right).
\label{Sfinal2D}
\end{equation}
Furthermore, from Eqs.~(\ref{KRRter}) and (\ref{SPgh2D}) one has also
the asymptotic relation
\begin{equation}
K^{Kr}_v\left(t_i,r_i;t_f,r_f;s\right)
\approx \sum_{\nu\in\Z} K_0
\left(t_i,r_i;t_f,r_f+i\beta\nu;s\right).
\label{Sfinal2D*}
\end{equation}

In conclusion, {\it the 2D Gui space is the approximate space for the
2D Schwarzschild black hole asymptotically far away from it, in the
sense of section \ref{sec:intro}}.

\subsection{Four dimensional case}

In the 4D case, the Schwarzschild black-hole line element is given by
\begin{equation}
ds^2_g = \frac{1}{\kappa^2}\,
\left[\,1-\frac{2M}{r}\,\right]\,\frac{du^+\,du^-}{u^+\,u^-}
- r^2\,d\Omega^2,
\label{Sds4D}
\end{equation}
where $\Omega$ is the solid angle, $\kappa=(4M)^{-1}$ and where $r$
and $t$ are also given by Eqs.~(\ref{u21}) and (\ref{u22}).  In a
similar way as that of the 2D case, one defines real and complex
tortoise coordinates.  One then finds that the covering space in
complex tortoise coordinates is given by
\begin{multline}
\Vct = \Biggl\{\,(\vt^0,\vt^1,\Omt)\in
\Biggr. \\[2mm] \left.
\left[\bigcup_{\nu\,\in\,\Z} A_{\nu/4} \times B_\nu \times\s^2\right] 
\bigcup \left[H_0\times H_1 \times \s^2\right]\mid r(v)>0\,\right\};
\label{Vct4d}
\end{multline}
see Eq.~(\ref{Vct2d}). Equations (\ref{KRR2d}) to (\ref{KRP2d})
obtained in the 2D case can be easily generalised to the 4D case. One
then has
\begin{align}
K^{Kr}_{_{RR}}\left(v_i;v_f;s\right) = \sum_{\nu\in\Z} 
K^B_v\left(v_i;v^0_f+i\beta\nu,v^1_f,\Omega_f; s\right),
\label{KRR4d} 
\end{align}
\vspace{-5mm}
\begin{multline}
K^{Kr}_{_{RF}}\left(v_i;v_f;s\right) = \\[2mm]
\sum_{\nu\in\Z} K^B_v \left(v_i;v^0_f+i\beta(\nu+1/4),
v^1_f-i\beta/4,\Omega_f; s\right),
\label{KRF4d}
\end{multline}
\vspace{-5mm}
\begin{align}
K^{Kr}_{_{RL}}\left(v_i;v_f;s\right) = 
\sum_{\nu\in\Z} K^B_v \left(v_i;v^0_f+i\beta(\nu+1/2),
v^1_f,\Omega_f; s\right),
\label{KRL4d}
\end{align}
\vspace{-8mm}
\begin{multline}
K^{Kr}_{_{RP}}\left(v_i;v_f; s\right) = \\[2mm]
\sum_{\nu\in\Z} K^B_v \left(v_i;v^0_f+i\beta(\nu+3/4),
v^1_f-i\beta/4,\Omega_f; s\right),
\label{KRP4d}
\end{multline}
and other results similar to those of Sec.~\ref{sec:pi}.

\begin{figure*}
\centerline{ \epsfysize=2.0truein \epsfbox{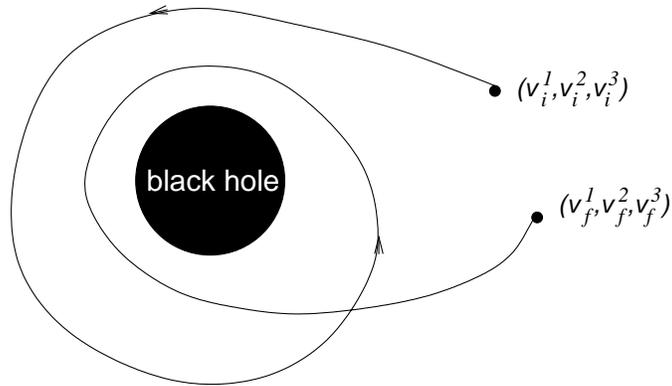}}
\mbox{\vspace{0mm}}
\caption{A path is shown in the space projection of a Schwarzschild
black-hole spacetime. The path winding number $\mu$ with respect to
the black-hole singularity is $+2$.}
\label{fig:4d}
\end{figure*}

\bigskip
In the 4D black-hole spacetime, there are a infinite number of
geodesics joining two end points, contrary to the 2D case.  This gives
a multiply connected structure to this spacetime. In its space
projection, the geodesics have a well defined winding number $\mu$
around the origin $r=0$; see Fig.~\ref{fig:4d}. The geodesics will not
cross the horizon if the end points are located outside the black
hole. In this case, one writes from Eq.~(\ref{GFK})
\begin{multline}
K^{Kr}_u \left(u_i;u_f;s\right) = \frac{1}{\left(4\pi\hbar
is\right)^2}\sqrt{\Delta(u_i;u_f)}\ F(u_i;u_f;s) 
\\[3mm] \times
\sum_{\mu\in\Z}\exp\left[\frac{i}{\hbar}
\frac{\sigma_{g_u}(u_i,u_f;\mu)}{4s}\right],
\label{ClAp}
\end{multline}
where $\sigma_{g_v}(v_i,v_f;\mu)$ is the proper arc length between the
end points along the geodesic of winding number $\mu$. Some of these
geodesics probe spacetime close to the black hole, where Gui's metric
for $D=4$, Eq.~(\ref{Gds}), and the 4D black-hole metric,
Eq.~(\ref{Sds4D}), differ significantly. Since these geodesics are not
contained entirely within the region $\Rc$, Eq.~(\ref{SPgh2D}),
obtained in the 2D case, is not true in the 4D case.  Physically, this
means that the potential barrier close to the black hole modifies the
properties of the radiation.  If its influence is neglected, as is it
often done in the literature when treating the 4D case,
Eq.~(\ref{SPgh2D}) will also be true in the 4D case under the
condition that the end points are both contained within a relatively
small solid angle, so that the forms of the 4D Gui and black-hole line
elements, Eq.~(\ref{Gds}) and Eq.~(\ref{Sds4D}), are the same. One
will then obtain
\begin{multline}
K^{Kr}_v\left(t_i,r_i,\Omega_i;t_f,r_f,\Omega_f;s\right) \\[2mm]
\approx \sum_{\nu\in\Z} K_0
\left(t_i,r_i,\Omega_i;t_f+i\beta\nu,r_f,\Omega_f; s\right).
\label{Sfinal}
\end{multline}
However, this last result does not give any information about the
asymptotic form of $K^{Kr}_v$. It can at most be considered as an
approximation, whose validity is difficult to estimate.

\subsection{Collapsing black holes}

The 2D and 4D collapsing Schwarzschild black holes are reviewed in
Ref.~\cite{OV}, where the method exposed in the present paper was
applied to these non-static spacetimes. I shall here only generalise
the results of Sec.~\ref{sec:UG} to these non-static black-hole
spacetimes.

\bigskip
The exterior and interior spacetime regions of the collapsing black
holes shall be denoted by $I$ and $I\!I$ respectively. By generalising
Eq.~(\ref{GfinalNS}), one has in the 4D case,
\begin{align}
K^{Kr}_v\left(v_i;v_f;s\right)
= \sum_{\nu\in\Z} K^B_v
\left(\vt_i;\vt^+_f,\vt^-_f+i\beta\nu,\Omt_f;s\right),
\label{GfinalNSbh}
\end{align}
where $\mid\!B\!>$ is a Boulware-like vacuum.  Defining the base-point
values as in Eqs.~(\ref{NSbaseI}) and (\ref{NSbaseII}), one deduces
from this last equation,
\begin{multline}
K^{Kr}_{_{I,I}} \left(v_i;v_f;s\right) =
K^{Kr}_{_{I\!I,I\!I}} \left(v_i;v_f;s\right) \\[2mm]
=\sum_{\nu\in\Z} 
K^B_v \left(v_i;v^+_f,v^-_f+i\beta\nu,\Omega_f;s\right),
\label{KIIbh}
\end{multline}
\vspace{-5mm}
\begin{multline}
K^{Kr}_{_{I,I\!I}} \left(v_i;v_f;s\right) \\[2mm] 
= \sum_{\nu\in\Z} K^B_v
\left(v_i;v^+_f,v^-_f+i\beta(\nu+1/2),\Omega_f;s\right);
\label{KIIIbh}
\end{multline}
see Eqs.~(\ref{KII}) and (\ref{KIII}). In particular, quantum
evolution is unitary in the fifth variable. Similar results are
obtained in the 2D black-hole case.

\section{Conclusions}

In this paper it was shown that it is possible to evaluate quantum
mechanical path integrals in spacetimes endowed with
event-horizons. In order to do so, we worked in tortoise coordinates,
for which the metric looks like the Minkowski metric, at least in the
region of interest, for example far away from the black hole.  We then
introduced complex tortoise coordinates to cover the entire spacetime
at once. The global properties of the path integral, which are related
to the boundary conditions of the propagator, have then been exploited
to obtain its thermal features, via the spacetime non-trivial
associated topology in complex tortoise coordinates.

\smallskip
An advantage of using complex tortoise coordinates relies on the fact
that some global issues can be addressed in these coordinates, such as
the calculation of the propagator whose end points are located in
spacetime regions separated by one or several event-horizons. It is
not clear to me that it is possible to do this within the framework of
a strict Euclidian approach.  The hermiticity property of the
propagator can then be analysed globally. In particular, it was shown
that quantum evolution in the fifth variable is unitary through the
event-horizons in both the static and non-static versions of the Gui
and Schwarzschild spacetimes.

\section*{Acknowledgements}

The author is very grateful to M.E.~Ortiz for numerous and enlighting
discussions, to M.~Blasone and M.~Zamora for precise and pertinent
remarks and to T.~Evans and D.~Steer for reading earlier versions of
the manuscript.  This work was supported by the Swiss National Science
Foundation.
 
\section*{Appendix}

In this Appendix, one redefines and compute the sums over paths in
Gui's spacetime.

\smallskip
In the Kruskal covering space $\Xct$, the sum over paths in
Eq.~(\ref{SPx}) is defined as a path integral whose integration over
the variable $\xt^-$ is performed by using the principal value,
i.e.~one defines
\begin{align}
\sideset{}{_{h_x}}\sum_{\substack{x_i\rightarrow x_f \\ [x]\,\in\,\Xc}}
\exp\left(\frac{i}{\hbar} S_{h_x}[x]\right) 
= \underset{[x]\,\in\,\Xct'}{\int_{\xt_i}^{\xt_f}} 
D_{\hti_\xt}[\xt]\, 
\exp\left(\frac{i}{\hbar} \widetilde{S}_{\hti_\xt}[\xt]\right),
\label{PIx*}
\end{align}
if the space $\Xct'$ is given by
\begin{align}
\Xct' = \{\,(\xt^+,\xt^-,\xft)\in\R\times\R'\times\R^n\,\},
\end{align} 
where $\R'=\lim_{\epsilon\rightarrow{0}}\,(-\infty,-\epsilon]
\cup[+\epsilon,+\infty)$.

\smallskip
In the covering space $\Yct$, since the sum over paths on the RHS of
Eq.~(\ref{SPxy}) is also badly defined when written as a path
integral, one introduces the space $\Yct'$ by
\begin{align}
\Yct' &= \{\,(\yt^0,\yt^1,\yft)\in
\bigcup_{\mu\,\in\,\Z} A_{\mu/4}\times B_\mu \times\R^n \},
\end{align}
i.e.~one removes from the covering space $\Yct$ the set parametrising
the ``event-horizon'' as in Kruskal coordinates. One then rewrites the
sum over paths in the form
\begin{equation}
\sideset{}{_\eta}\sum_{\substack{y_i\rightarrow y_f \\ [y]\,\in\,\Yc}}
\exp\left(\frac{i}{\hbar} S_\eta[y]\right) = 
\sum_{\nu\in\Z} \underset{[\yt]\,\in\,\Yct'}{\int_{\yt_i}^{\yt^\nu_f}} 
D_\etat[\yt]\,\exp\left(\frac{i}{\hbar} \widetilde{S}_\etat[\yt]\right),
\label{pii}
\end{equation}
where $\etat=\eta$. In this last equation, we have taken into account
the multiply connected nature of $\Yc$ by summing over the homotopic
classes of winding number $\nu$.

\smallskip
In Eq.~(\ref{pii}) each path integral in the sum contains the term
\begin{multline}
\int_{\Yct'} \!\! d^D\yt_j \
\exp\left(\frac{i}{4\hbar} \sum_{k=1}^N\int_{s_{k-1}}^{s_k}\!
d\omega \, \widetilde{L}_g[\yt,\dot{\yt}] \right)
\\[2mm] =
\sum_{\mu\in\Z} \int_{A_{\mu/4}}\!\!\!\!d\yt^0_j\int_{B_\mu}\!\!\!d\yt^1_j
\int_{\R^n}\!\!\! d\yft_j \qquad\qquad\qquad\qquad
\\[2mm] \times
\exp\left(\frac{i}{4\hbar} \sum_{k=1}^N\int_{s_{k-1}}^{s_k} \!
d\omega\,\etat_{\mu\nu}\,\dot{\yt}^\mu(\omega)\,\dot{\yt}^\nu(\omega)\right).
\end{multline}
Since the integrand in the RHS of this last equation does not
actually depend on the imaginary value of $\yt^-_j$, one has
\begin{multline}
\int_{\Yct'} d^D\yt_j \ \exp\left(\frac{i}{4\hbar} \
\sum_{k=1}^N\int_{s_{k-1}}^{s_k}\! d\omega \, 
\etat_{\mu\nu}\,\dot{\yt}^\mu(\omega)\,\dot{\yt}^\nu(\omega)\right)
\\[2mm] = C \int_{\R^D} d^D\yt_j\
\exp\left(\frac{i}{4\hbar} \sum_{k=1}^N\int_{s_{k-1}}^{s_k}\!d\omega\, 
\etat_{\mu\nu}\,\dot{\yt}^\mu(\omega)\,\dot{\yt}^\nu(\omega)\right),
\end{multline}
where $C$ is an infinite constant. This constant is removed by
renormalising the path integral to take into account the fact that the
integration is performed over an infinite number of copies of $\R^D$.
The contribution of the homotopic class of winding number $\nu$ is
then the free propagator $\Kt_0$ with arguments $\yt_i$ and
$\yt_f^\nu$,
\begin{equation}
\underset{[\yt]\,\in\,\Yct'}{\int_{\yt_i}^{\yt^\nu_f}}
D_\etat[\yt]\,\exp\left(\frac{i}{\hbar} \widetilde{S}_\etat[\yt]\right) 
= \Kt_0\left(\yt_i;\yt_f^\nu; s\right).
\label{clnu}
\end{equation}
Equation (\ref{Gfinal}) is finally obtained from Eqs.~(\ref{pii}) and
(\ref{clnu}).


\newpage

\begin{table*}
\centerline{
\epsfysize=4.0truein
\epsfbox{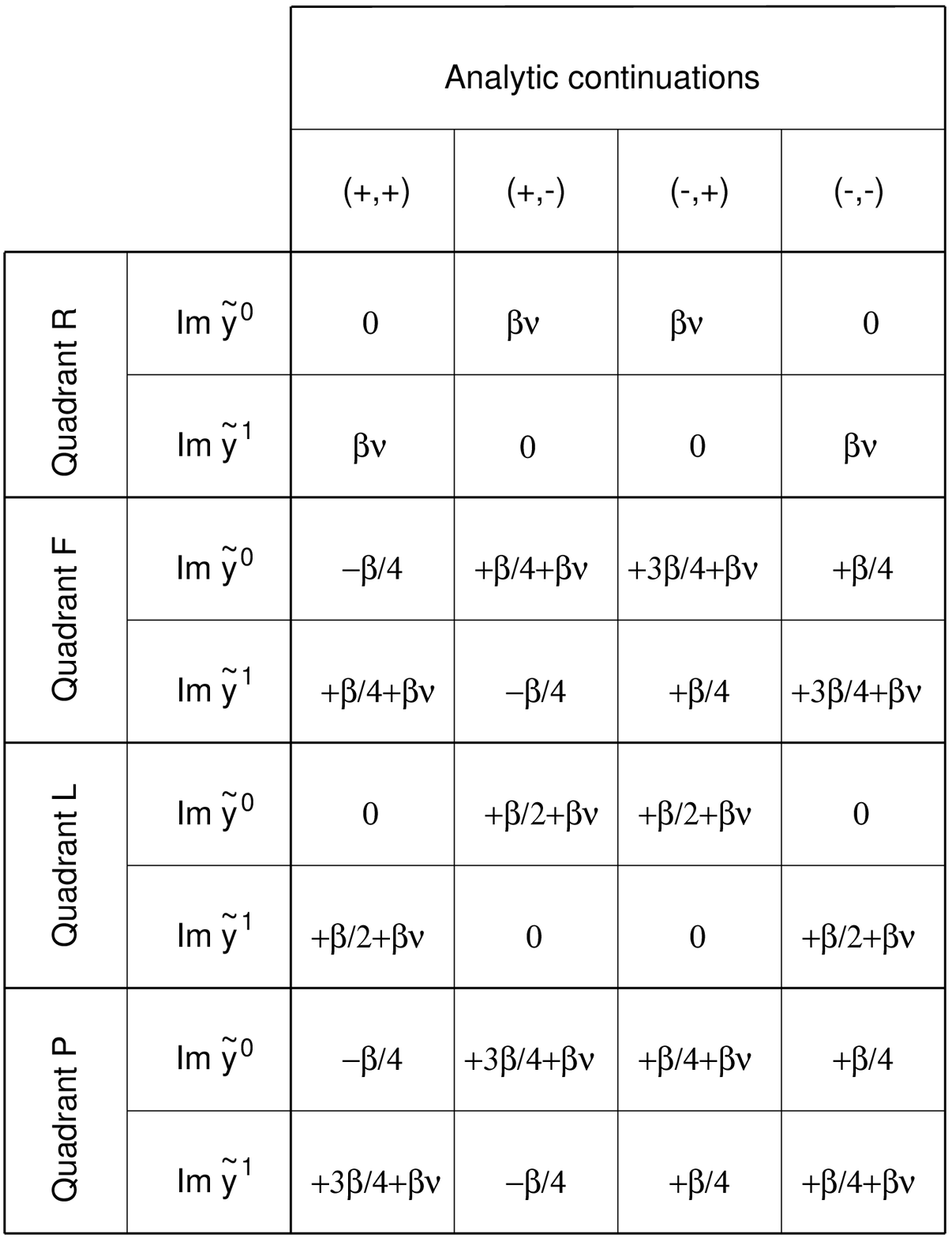}}
\mbox{\vspace{0mm}}
\caption{The four possible ways to continue analytically the functions
$\yt^0=\yt^0(\xt^0,\xt^1)$ and $\yt^1=\yt^1(\xt^0,\xt^1)$ in the
complex tortoise covering space ($\beta=2\pi/\kappa$, $\nu\in\Z$); see
Eq.~(\ref{recip}). The couple of signs on the first line indicates the
directions (anticlockwise ($+$) or clockwise ($-$)) with respect to
which the corresponding analytical extensions has been performed in
the $\xt^+$ and $\xt^-$ complex planes respectively assuming the
sequence $(R,F,P,L)$.}
\label{fig:param}
\end{table*}


\begin{thebibliography}{}


\bibitem{FHS} R.~P.~Feynman and A.~R.~Hibbs, {\it Quantum mechanics and
path integrals} (McGraw-Hill, New York, 1965).

\bibitem{Ma} \Ref{M.~S.~Marinov}{Phys.~Rep.}{60}{1}{1980}.

\bibitem{Sch1} L.~S.~Schulman, {\it Techniques and application of path
integrals} (Wiley, New York, 1981).

\bibitem{KL} \Ref{D.~C.~Khandekar and S.~V.~Lawande}
{Phys.~Rep.}{137}{115}{1986}. 

\bibitem{BP} \Ref{J.~D.~Bekenstein and L.~Parker}
{\PRD}{23}{2850}{1981}; L.~Parker, in {\it Recent
developments in gravitation, Carg\`ese 1978}, edited by M.~L\'evy and
S.~Deser (Plenum, New York, 1979).

\bibitem{HH} \Ref{J.~B.~Hartle and S.~W.~Hawking}{\PRD}{13}{2188}{1976}.

\bibitem{TV} \Ref{W.~Troost and H.~Van Dam}{Phys.~Lett.~B}{71}{149}{1977};
\Refa{Nucl.~Phys.~B}{152}{442}{1979}.

\bibitem{Co} \Ref{P.~Candelas and D.~J.~Raine}{\PRD}{15}{1494}{1977};
\Ref{D.~M.~Chitre and J.~B.~Hartle}{\PRD}{16}{251}{1977};
\Ref{I.~H.~Duru and N.~\"Unal}{\PRD}{34}{959}{1986}.

\bibitem{Be} \Ref{C.~C.~Bernido and G.~Aguarte}{\PRD}{56}{2445}{1997}.

\bibitem{Me} \Ref{M.~Mensky}{Theor.~and Math.~Phys.}{115}{542}{1998}.

\bibitem{OV} \Ref{M.~E.~Ortiz and F.~Vendrell}{\PRD}{59}{084005}{1999}.

\bibitem{Ha} \Ref{S.~W.~Hawking}{Commun.~Math.~Phys.~}{43}{199}{1975}.

\bibitem{BD} \Ref{P.~C.~W.~Davies}{Rep.~Prog.~Phys.}{41}{1313}{1978};
N.~D.~Birrell and P.~C.~W.~Davies, {\it Quantum fields in curved space},
Cambridge University Press, 1982.

\bibitem{Gui1} \Ref{Y.-X.~Gui}{Sci.~Sin.}{31A}{1104}{1988}.

\bibitem{Gui2} \Ref{Y.-X.~Gui}{\PRD}{42}{1988}{1990};
\Refb{46}{1869}{1992}; 
\Ref{Z.~Zhao and Y.-X.~Gui}{Nuov.~Cim.}{109B}{355}{1994}.

\bibitem{Isham} \Ref{B.~S.~DeWitt, C.~F.~Hart and C.~J.~Isham}
{Physica A}{96}{197}{1979}.

\bibitem{Sch2} \Ref{L.~S.~Schulman}{Phys.~Rev.}{176}{1558}{1968};
\Refa{J.~Math.~Phys.}{12}{304}{1971}.

\bibitem{AB} \Ref{Y.~Aharanov and D.~Bohm}{\PRD}{115}{485}{1959}.

\bibitem{Pol} \Ref{S.~F.~Edwards}{Proc.~Phys.~Soc.}{91}{513}{1967}.

\bibitem{AA} A.~Anderson, {\it private communication}.

\bibitem{BM} \Ref{M.~V.~Berry and K.~E.~Mount}{Rep. Prog. Phys.}
{35}{315}{1972}.

\bibitem{LF} \Ref{R.~Laflamme}{Nucl.~Phys.}{B324}{233}{1989}.

\bibitem{Sy} \Ref{J.~L.~Synge}
{Proc.~R.~Ir.~Acad.~Sect.~A, Math.~Astron.~Phys.~Sci.}{59}{1}{1957}. 

\bibitem{HO} \Ref{J.~J.~Halliwell and M.~E.~Ortiz}
{Phys.~Rev.~D}{48}{748}{1993}.

\bibitem{Um} H.~Umezawa, H.~Matsumoto and M.~Tachiki, 
{\it Thermo-field dynamics and condensed states} (North-Holland, 
Amsterdam, 1982).

\bibitem{LL} \Ref{M.~Lachi\`eze-Rey and J.-P.~Luminet}
{Phys.~Rep.}{254}{136}{1995}; J.-P.~Luminet and B.~F.~Roukema, {\it
Proceedings of Cosmology School, Cargese 1998}.

\bibitem{Do} \Ref{J.~S.~Dowker}{J.~Phys.~A}{5}{936}{1972}.

\end{thebibliography}
\end{document}